\documentclass[aps,prc,reprint,groupedaddress]{revtex4-1} 
\usepackage{amsmath}
\usepackage{graphicx}
\usepackage{array}
\usepackage{capt-of}

\begin{document}
\newcommand{\PIEL}{$\pi N \rightarrow \pi N$}	
\newcommand{\GPPN}{$\gamma N \rightarrow \pi N$}
\newcommand{\GPEP}{$\gamma p \rightarrow \eta p$}
\newcommand{\GPKL}{$\gamma p \rightarrow K \Lambda$}
\newcommand{\GNEN}{$\gamma n \rightarrow \eta n$}
\newcommand{\PINE}{$\pi^- p \rightarrow \eta n$}
\newcommand{\PINK}{$\pi^- p \rightarrow K^0 \Lambda$}
\newcommand{\DSG}{$d\sigma / d \Omega$}

\newcommand{\trimAA}{3mm}
\newcommand{\trimBB}{10mm}
\newcommand{\trimCC}{15mm}
\newcommand{\trimDD}{19mm}

\newcommand{\pathGPEP}{Figures/GPEP/}
\newcommand{\pathGNEN}{Figures/GNEN/}
\newcommand{\DSGT}{$\frac{d\sigma}{d\Omega}$}

\newcommand{\HSP}{H_{SP}}
\newcommand{\HSA}{H_{SA}}
\newcommand{\HN}{H_{N}}
\newcommand{\HD}{H_{D}}
\newcommand{\HSPC}{H_{SP}^*}
\newcommand{\HSAC}{H_{SA}^*}
\newcommand{\HNC}{H_{N}^*}
\newcommand{\HDC}{H_{D}^*}
\newcommand{\qk}{\frac{q}{k}}
\newcolumntype{C}{>{ \arraybackslash}p{7em}}

\newcommand{\bibA}[4]{#1, #2 \textbf{#3}, #4}
\newcommand{\bibE}[4]{#1 \textit{et al.}, #2 \textbf{#3}, #4.}
\newcommand{\PRL}{Phys. Rev. Lett.}

\newcommand{\printObs}{\DSGT}
\newcommand{\printReac}{\GPEP}
\newcommand{\ObsName}{DSG}
\newcommand{\pathGraphing}{\pathGPEP}

\newcommand{\trimA}{6mm}
\newcommand{\trimB}{4mm}
\newcommand{\trimC}{10mm}
\newcommand{\trimD}{22mm}
\newcommand{\figWidth}{.30}


\title{Partial-wave analyses of $\gamma p \rightarrow \eta p$ \ and $\gamma n \rightarrow \eta n$ using a multichannel framework}


	\author{B. C. Hunt and D. M. Manley}
	\affiliation{Department of Physics, Kent State University, Kent, OH 44242-0001, USA}
	

\date{\today}

\begin{abstract}
	This paper presents results from partial-wave analyses of the photoproduction reactions $\gamma p \rightarrow \eta p$  and $\gamma n \rightarrow \eta n$. World data for the observables \DSG, $\Sigma$, $T$, $P$, $F$, and $E$ were analyzed as part of this work. The dominant amplitude in the fitting range from threshold to a c.m.\ energy of 1900~MeV was found to be $S_{11}$ in both reactions, consistent with results of other groups. At c.m.\ energies above 1600 MeV, our solution deviates from published results, with this work finding higher-order partial waves becoming significant. Data off the proton suggest that the higher-order terms contributing to the reaction include $P_{11}$, $P_{13}$, and $F_{15}$. The final results also hint that $F_{17}$ is needed to fit double-polarization observables above 1900~MeV. Data off the neutron show a contribution from $P_{13}$, as well as strong contributions from $D_{13}$ and $D_{15}$. 
\end{abstract}


\maketitle

\section{Introduction}
In recent years, a wealth of new high-precision experimental data has been measured at various facilities including JLab, MAMI, LEPS, SLAC, and GRAAL for a number of observables with the goal of better understanding the spectrum of $N^*$ and $\Delta^*$ resonances. Despite past efforts, there are still predicted resonances that have not been found, known as the problem of the ``missing resonances'', and other resonances whose properties are not well determined. Two possible explanations for this are that (1) the missing resonances do not exist or (2) they couple mainly to reactions not yet analyzed. This work investigates the second possibility. Knowledge gained from this and future work is expected to guide theorists trying to understand the fundamental features of Quantum Chromodynamics (QCD) or the theory of the quarks and gluons that bind matter into hadrons. 

It has been shown that at least eight measured observables are needed to perform a complete experiment \cite{Tabakin}. The database analyzed in this work for $\gamma p \rightarrow \eta p$ and $\gamma n \rightarrow \eta n$ include significant amounts of data for five of the eight needed observables. These five are \DSG, $\Sigma$, $T$, $F$, and $E$ measured at various c.m.\ energies from threshold to 1900~MeV. Also analyzed were seven $P$ and 12 $C_x$ data points. Data for the helicity-dependent cross section were analyzed as \DSG \ and $E$ data. Table \ref{tbl:DatabaseCount} tabulates the number of data available for each observable and shows that while there are a wealth of differential cross-section data, the polarization measurements are still limited. Because there are still insufficient data for a complete experiment, information from other reactions, including $\gamma N \rightarrow \pi N$ and $\pi N \rightarrow \eta N$, was used to constrain the fits.
\begin{table}[bpht]

	\renewcommand{\columnwidth}{4.5}
	\begin{tabular}{ccccc}
		\hline\hline  Observable  & \GPEP  & References &  \GNEN  & References \\ \hline
		 \DSG         & 7754 & \cite{Heusch66,Delcourt69,Christ73,Booth74,Vartapetyan80,Homma88,Dytman95,Krusche95,Dugger02,Ahrens03,Crede05,Nakbayashi06,Bartalini07,Crede09,Sumihama09,Williams09,Mcnicoll10,Kashevarov17} & 879  & \cite{Werthmuller} \\
		  $T$         & 439  & \cite{Akondi14,Krusche15} & 96   & \cite{Krusche15} \\
		  $\Sigma$    & 236  & \cite{Vartapetyan80,Ajaka98,Kouznetsov98,Bartalini07,Elsner07} & 80   & \cite{Fantini} \\  
		  $P$         & 7    & \cite{Heusch70,Hongoh71,Sarantsev} &  0   &  \\
		  $E$         & 331  & \cite{Senderovich14,Witthauer17} & 135  & \cite{Witthauer17} \\ 
		  $F$         & 241  & \cite{Akondi14,Krusche15} & 96   &\cite{Krusche15}  \\ 
		 $C_x$ & 12  & \cite{Sikora10} &  0 \\ 
		\hline\hline
	\end{tabular} 
	\caption{\label{tbl:DatabaseCount}  Number of  experimental data used in our analysis for each fitted observable.  Preliminary data from Ref.~\cite{Krusche15} were included in the analysis but are not shown in the figures.} 
\end{table}

This work analyzed the world data of eta photoproduction off the nucleon in the c.m.\ energy range from threshold up to almost 2000~MeV. The final generated energy-dependent solutions were then used in the KSU multichannel framework to improve knowledge about the $N^*$ and $\Delta^*$ resonance parameters. Section~II outlines the basic formalism used throughout this work including sign conventions for the different spin observables.  Section~III describes the general procedure that we used to obtain the results. Section~IV describes results of the analyses for the reactions $\gamma p \rightarrow \eta p$ and \GNEN. Comparisons to results from BnGa (2016) \cite{Sarantsev} and J\"ulich (2015) \cite{Julich} are also shown. Fits to the data are shown in Appendix~A.

\section{Formalism}
Four helicity amplitudes are needed to describe the photoproduction of a pseudoscalar ($J^P=0^-$) meson and a $J^P=\frac{1}{2}^+$ baryon off of a nucleon target \cite{BDS1}. Each of the four helicity amplitudes can be expanded in terms of electric and magnetic multipoles $E_{l \pm}$ and $M_{l\pm}$, respectively, where $l = 0, 1, 2, \ldots$ is the orbital angular momentum of the final-state hadrons and $j=l \pm \frac{1}{2}$ is the total angular momentum. Each multipole is a complex function of energy, which makes the helicity amplitudes complex functions of both energy and scattering angle:

	\begin{subequations}
	\begin{multline}
	\label{eq:hnfull}
	\HN = \sqrt{\frac{1}{2}} \cos\left(\frac{\theta}{2}\right) \sum_{l=0}^{\infty}\left[
	\left(l+2\right)E_{l+}
	+lM_{l+} \right.  \\
	+lE_{\left(l+1\right)-}
	-\left(l+2\right)M_{\left(l+1\right)-}\left. \right]\left(P_l^{'} -P_{l+1}^{'}\right) ,
	\end{multline}

	\begin{multline}
	\label{eq:hspfull}
	\HSP= \sqrt{\frac{1}{2}} \cos\left(\frac{\theta}{2}\right) \sin(\theta) \sum_{l=1}^{\infty}\left[
	E_{l+}-M_{l+} \right]\\
	-E_{\left(l+1\right)-}
	-M_{\left(l+1\right)-}\left. \right]\left(P_l^{''}-P_{l+1}^{''}\right),
	\end{multline}
	
	\begin{multline}	
	\label{eq:hsafull}
	\HSA = \sqrt{\frac{1}{2}} \sin\left(\frac{\theta}{2}\right) \sum_{l=0}^{\infty}\left[
	\left(l+2\right)E_{l+}\right.
	+lM_{l+} \\
	-lE_{\left(l+1\right)-}
	+\left(l+2\right)M_{\left(l+1\right)-}\left. \right]\left(P_l^{'} +P_{l+1}^{'} \right), 
	\end{multline}
	\begin{multline}	
	\label{eq:hdfull}
	\HD = \sqrt{\frac{1}{2}} \sin\left(\frac{\theta}{2}\right) \sin(\theta) \sum_{l=1}^{\infty}\left[
	E_{l+}-M_{l+} \right. \\
	+E_{\left(l+1\right)-}
	+M_{\left(l+1\right)-}\left. \right]\left(P_l^{''}+P_{l+1}^{''} \right). 
	\end{multline}	

	\label{eq:Helicity_Amplitudes}
\end{subequations}
The naming convention for the four helicity amplitudes above follows that of the SAID group \cite{SAID1990}. All 16 single- and double-polarization observables can be written in terms of these four helicity amplitudes; however, in the literature, different sign conventions are used in their definitions. The definitions for each of the observables included in this work are given in Table \ref{tbl:Observables}. 

\begin{table}
	\renewcommand{\arraystretch}{1.5}
	 
	\begin{tabular}{c >{ \arraybackslash}p{1.8cm}}\hline\hline
               Observable & Expt. \\
		\hline 
		$ \sigma(\theta) = \frac{q}{2k}\left[|\HN|^2 + |\HD|^2 + |\HSA|^2 + |\HSP|^2\right]$  & \\
		$ \Sigma \: \sigma(\theta) = \qk \text{Re}\left[\HSP\HSAC-\HN\HDC\right]$ & $B$\\
		$T  \: \sigma\left(\theta\right) = \qk \text{Im}\left[\HSP \HNC + \HD \HSAC\right]$&$T$ \\
		$P  \: \sigma\left(\theta\right) = - \qk \text{Im}\left[\HSP \HDC + \HN \HSAC\right]$ & $R$\\ 
		$G  \: \sigma\left(\theta\right) = -\qk \text{Im}\left[\HSP \HSAC + \HN \HDC\right]$& $B$, $T$ \\
		$H  \: \sigma\left(\theta\right) = -\qk \text{Im}\left[\HSP \HDC + \HSA \HNC\right]$& $B$, $T$ \\
		$F  \: \sigma\left(\theta\right) = \qk \text{Re} \left[\HSA\HDC + \HSP\HNC\right]$& $B$, $T$\\
		$E  \: \sigma\left(\theta\right) = \frac{q}{2k} \left[|\HN|^2 + |\HSA|^2 - |\HD|^2 - |\HSP|^2\right]$& $B$, $T$\\ \hline\hline
	\end{tabular}
	\caption{List of single-polarization and double-polarization observables analyzed in this work. See Refs.  \cite{BDS1, Sandorfi} for a detailed description of the necessary experimental setup and equations for all 16 observables. In the second column, $B$, $T$, and $R$ refer to a measurement of the beam, target, and recoil nucleon polarization, respectively. Note that $\sigma\left(\theta\right) = d\sigma/{d\Omega}$ is the differential cross section. }
	\label{tbl:Observables}
\end{table}

The literature also mentions measurements of $(d\sigma/d\Omega)_{\frac{1}{2}}$ and $(d\sigma/d\Omega)_{\frac{3}{2}}$, which are the helicity-dependent cross sections \cite{Witthauer17}. They are related to the $d\sigma/d\Omega$ and $E$ observables by
\begin{equation}
\frac{d\sigma}{d\Omega}_{\frac{1}{2}} = \frac{d\sigma}{d\Omega} + E\frac{d\sigma}{d\Omega} \propto |\HN|^2 + |\HSA|^2 
\label{eqH12}
\end{equation}
and
\begin{equation}
\label{eqH32} 
\frac{d\sigma}{d\Omega}_{\frac{3}{2}} = \frac{d\sigma}{d\Omega} - E\frac{d\sigma}{d\Omega} \propto |\HSP|^2 + |\HD|^2.
\end{equation}
$H_N$ and $H_{SA}$ are then pure helicity-$1/2$ amplitudes, while $H_{SP}$ and $H_D$ are pure helicity-$3/2$ amplitudes. The full differential cross section is recovered by the relationship
\begin{equation}
\frac{d\sigma}{d\Omega} = \frac{1}{2}\left[\frac{d\sigma}{d\Omega}_{\frac{1}{2}} + \frac{d\sigma}{d\Omega}_{\frac{3}{2}}\right].
\end{equation}
Equations \ref{eqH12} and \ref{eqH32} can be separately integrated to obtain what are called the helicity-1/2 and helicity-3/2 cross sections, which are shown in Figs.~\ref{SGTGPEP12} and \ref{SGTGPEP32}.

\section{Fitting Procedure}
We began our analysis by performing an independent single-energy partial-wave analyses of \GPEP. In this approach, partial-wave amplitudes are determined before adding information from a previously determined resonance structure. This achieved the goal of limiting bias in the single-energy partial-wave amplitudes at the beginning of the analysis. Only after an initial determination of the amplitudes was made were model constraints added to maintain consistency with other hadronic and photoproduction reactions. The starting point for our $\gamma n \rightarrow \eta n$ solution used amplitudes predicted from a multichannel fit determined after the $\gamma p \rightarrow \eta p$ analysis was in its final stages. This procedure was used because of the limited availability of $\gamma n \rightarrow \eta n$ data, as well as the relatively late stage when analysis of this reaction was first considered.

The starting point was to assemble all data within specified small c.m.\ energy ranges into individual bins. Observables within a single bin were then approximated as functions of just the scattering angle. It was determined that 5-MeV wide bins were needed near threshold where the $S_{11}$ amplitude dominates due to the rapid rise in the cross section near the $S_{11}$(1535) resonance. At c.m.\ energies above $W \approx 1600$~MeV, a trade-off between small bin sizes and keeping sufficient polarization data within the energy bin meant that larger bin sizes of 15 to 20~MeV were needed to constrain the fits. 

A known concern in performing a single-energy fit is that of the continuum ambiguity \cite{Svarc2018}, which permits a global change in phase to all of the partial-wave amplitudes with no observable change in the data. To address this ambiguity, the data $\gamma p \rightarrow \eta p$ were initially fitted with a purely real $S_{11}$ amplitude to determine its magnitude. Then an energy-dependent fit of several $S_{11}$ amplitudes, similar to those of Shrestha and Manley \cite{ShresthaED}, was used to determine its phase through unitarity constraints.  The energy-dependent fits included available single-energy amplitudes for individual partial waves from the $\gamma N \rightarrow \pi N$, $\gamma p \rightarrow \eta p$, $\gamma n \rightarrow \eta n$, $\gamma p \rightarrow K^+ \Lambda$, $\pi N \rightarrow \pi N$, $\pi N \rightarrow \pi \pi N$, $\pi N \rightarrow \eta N$, and $\pi N \rightarrow K \Lambda$ reactions.  With the $S_{11}$ amplitude for $\gamma p \rightarrow \eta p$ fully determined, initial values for the higher-order amplitudes could then be determined. For \GNEN, the phase and magnitude of each partial wave were initially determined from the energy-dependent fit.

Due to complexities that arise from interference effects, an iterative procedure was needed to obtain good quality fits to the data. The procedure involved two main steps that were iterated as many times as necessary to obtain convergence. The first step (single-energy fits) was to allow a subset of the partial-wave amplitudes (including $S_{11}$ as needed) to vary in each energy bin. This generated a discrete solution for each of the varied partial-wave amplitudes. These single-energy results were then used as input to energy-dependent fits (the second step) that were also used to determine the resonance parameters. In this second step, resonance parameters were adjusted to generate a smooth energy-dependent solution of the single-energy amplitudes. Finally, the output of the second step was used as input to the first step. This iterative procedure was continued until $\chi^2$ reached a global minimum for this and all other analyzed reactions in the energy-dependent analysis.  

The single-energy fits described above used a modified gradient descent algorithm \cite{Bevington} to determine an optimal set of values for the partial-wave amplitudes, with each bin's parameters being treated as independent. Because not all measured observables are available in all energy bins, the algorithm's standard $\chi^2$ function allowed too much variation in the amplitudes between different energy bins. A penalty term was added to the standard $\chi^2$ term to limit this bin-to-bin variation in the solution. This had the desired effect of improving the fits at the expense of permitting only small updates to the parameters during each iteration. The explicit form for a penalty term was
\begin{equation}
\chi^2_{\text{penalty}} = f [(PW^R_{\text{ED}} - PW^R_{\text{fit}})^2 + (PW^I_{\text{ED}} - PW^I_{\text{fit}})^2],
\end{equation}
where $PW^R_{\text{ED}}$ and $PW^I_{\text{ED}}$ are the real and imaginary parts of the partial-wave amplitude found in the preceding energy-dependent fit and $PW^R_{\text{fit}}$ and $PW^I_{\text{fit}}$ are the corresponding real and imaginary parts of the amplitude determined during each step of the single-energy fit. The factor $f$ was a parameter chosen to control the strength of the penalty term.  For the initial round of single-energy fits, we set $f=0$ for no penalty term at all. After the first round of energy-dependent fits, we used values from the energy-dependent fits to constrain selected partial waves in the next round of single-energy fits. This was initially done with a weak penalty constraint (e.g., $f=10$), but as iterations progressed and the energy-dependent solutions did a better job of describing the fitted observables, the strength of the penalty term was adjusted up to $f=100$. This biased results to single-energy solutions that were somewhat similar to the current energy-dependent solution. To verify the penalty term wasn't causing the fits to converge to a local minimum, multiple starting solutions were used to determine which potential solution produced the best fit. 

Once the single-energy solutions and energy-dependent solutions converged to both give a good description of the observables, final uncertainties on the amplitudes in the single-energy solutions from step one were obtained by fixing the phases of each partial-wave amplitude at the values from our energy-dependent solution, and then allowing only their moduli to vary. This phase constraint was needed to fix both the global phase of the solution and to constrain the final results due to lack of all spin observables at all energies. An additional penalty constraint was used as well, but kept small enough that the penalty contribution to $\chi^2$ was less than 10$\%$.  The resulting single-energy solutions, projected into real and imaginary parts, were then used as input to a final energy-dependent fit in which all parameters were free to vary, to determine final uncertainties in the $N^*$ and $\Delta^*$ resonance parameters.  Our final fits included all amplitudes up to $G_{17}$ for both $\gamma p \rightarrow \eta p$ and $\gamma n \rightarrow \eta n$, although $F_{17}$  was the highest amplitude necessary for good fits. 

\section{Results}
This section presents final results for the partial-wave analyses of both \GPEP \ and \GNEN. It compares results with those of other groups and shows the quality of agreement for the integrated cross-section data that were not directly fitted.

\subsection{\GPEP}
For the reaction \GPEP, the fits of the observables \DSG, $T$, $F$, and $E$ were very good over the entire energy range; however, the fits of the beam asymmetry $\Sigma$ showed minor problems at backward angles in the c.m.\ energy range 1650 to 1800 MeV. Table \ref{tab:BnGavsKSU} shows the $\chi^2$ contribution from each individual observable for the different works as well as the total over all observables. We note that the $\chi^2$ contribution from the differential cross section obtained in this work is significantly smaller than that by other groups with minimal impact to the spin observables. This is in part because new high-precision data for some observables used in this work were unavailable to the other groups at their time of analyses. In order to provide a good description of the data, the amplitudes $S_{11}$ and $D_{13}$ were needed starting at threshold and $P_{11}$, $P_{13}$, and $F_{15}$ waves were important above 1600 MeV. At energies above $1900$~MeV, $F_{17}$ appears important, but the lack of spin observables at these energies made it difficult to make any definitive conclusions. The magnitude of $S_{11}$ partial-wave amplitude above 1600~MeV was found to differ from the results of BnGa \cite{Sarantsev}, which found $S_{11}$ saturating the integrated cross section over almost the entire energy range analyzed in this work.

\begin{table}[bpht]
	 
	\renewcommand{\arraystretch}{1.5}
	\begin{tabular}{cccc}
		\hline\hline
		Observable	& KSU & BnGa (2016) & J\"ulich (2015b) \\ \hline
		\DSG	& 44000 & 83000 & 58000   \\
		$T$    	& 1500 & 1200 & 900   \\
		$\Sigma$& 950 & 380 & 1100    \\
		$F$	    & 680 & 480 & 340    \\
		$E$	    & 620 & 690 & 1300  \\
		$C_x$	& 16 & 20 & 30    \\
		$(d\sigma/d\Omega)_{\frac{1}{2}}$	  & 810 & 550 & 1300 \\
		$(d\sigma/d\Omega)_{\frac{3}{2}}$	  & 200 & 250 & 350  \\
		\hline 
		Fit Total& 50000 & 87000 & 64000  \\
		\hline\hline
	\end{tabular}
	\caption{\label{tab:BnGavsKSU} $\chi^2$ contributions for $\gamma p \rightarrow \eta p$. Column~1 provides the observable name, column~2 the $\chi^2$ contribution from this work, column~3 the $\chi^2$ contribution from BnGa (2016) \cite{Sarantsev} and column~4 the $\chi^2$ contribution from J\"ulich (2015b) \cite{Julich}. $\chi^2$ values, were obtained by binning the data in 5-MeV increments in the full c.m.\ energy range from 1490 to 1975~MeV.}
				
\end{table}

\begin{figure}[htp]
	
	\includegraphics[scale=.99,trim={20mm} {190mm} {100mm} {\trimDD},clip=true]{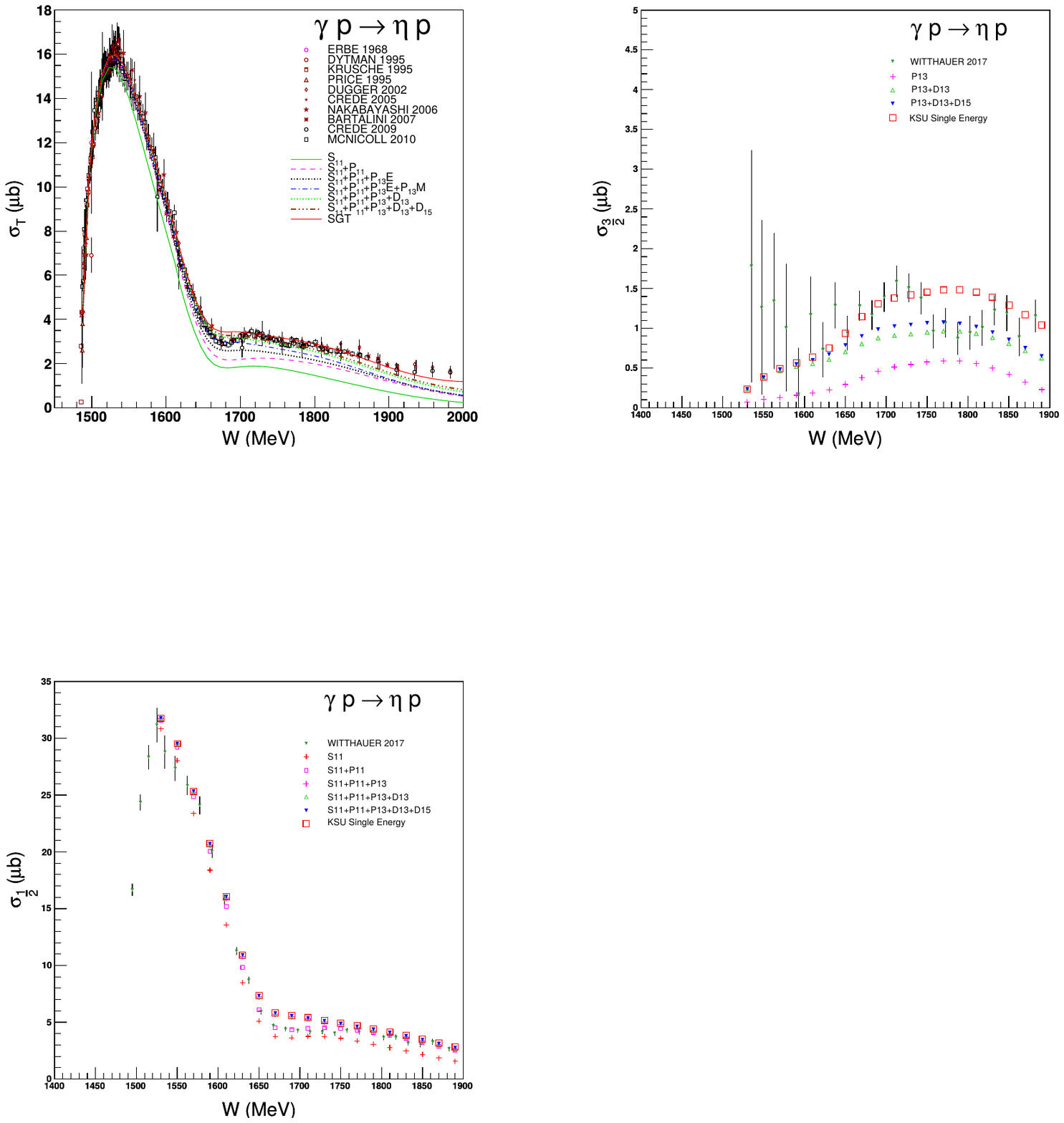}
	\caption[Integrated cross section for \GPEP.]{Integrated cross section for \GPEP. The data points are from ERBE 1968 \cite{Erbe68}, DYTMAN 1995 \cite{Dytman95}, KRUSCHE 1995 \cite{Krusche95}, PRICE 1995 \cite{Price95}, DUGGER 2002 \cite{Dugger02}, CREDE 2005 \cite{Crede05}, NAKABAYASHI 2006 \cite{Nakbayashi06}, BARTALINI 2007 \cite{Bartalini07}, CREDE 2009 \cite{Crede09}, and MCNICOLL 2010 \cite{Mcnicoll10}. The curves also show the contribution to the cross section by successively adding each partial wave.}
	\label{SGTGPEP}
\end{figure}

\begin{figure}[htp]

	\includegraphics[scale=.99,trim={22mm} {82mm} {110mm} {124mm},clip=true]{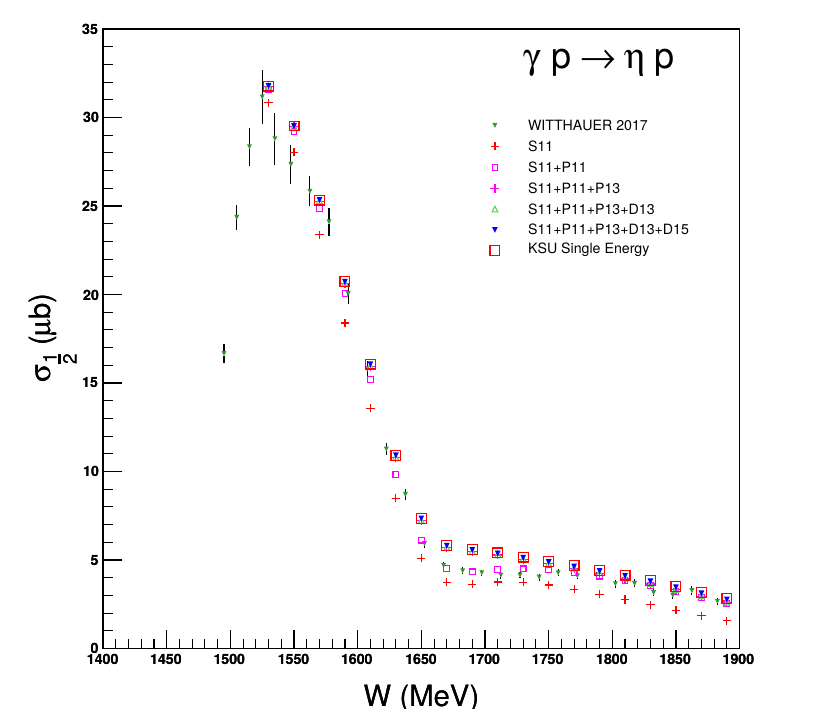}
	\caption{Helicity-$1/2$ integrated cross section for \GPEP. The data points are from Witthauer 2017 \cite{Witthauer17}. The plot also shows the contribution to the cross section by successively adding each partial wave.}
	\label{SGTGPEP12}
\end{figure}

\begin{figure}[htp]

	\includegraphics[scale=.99,trim={113mm} {188mm} {\trimCC} {\trimDD},clip=true]{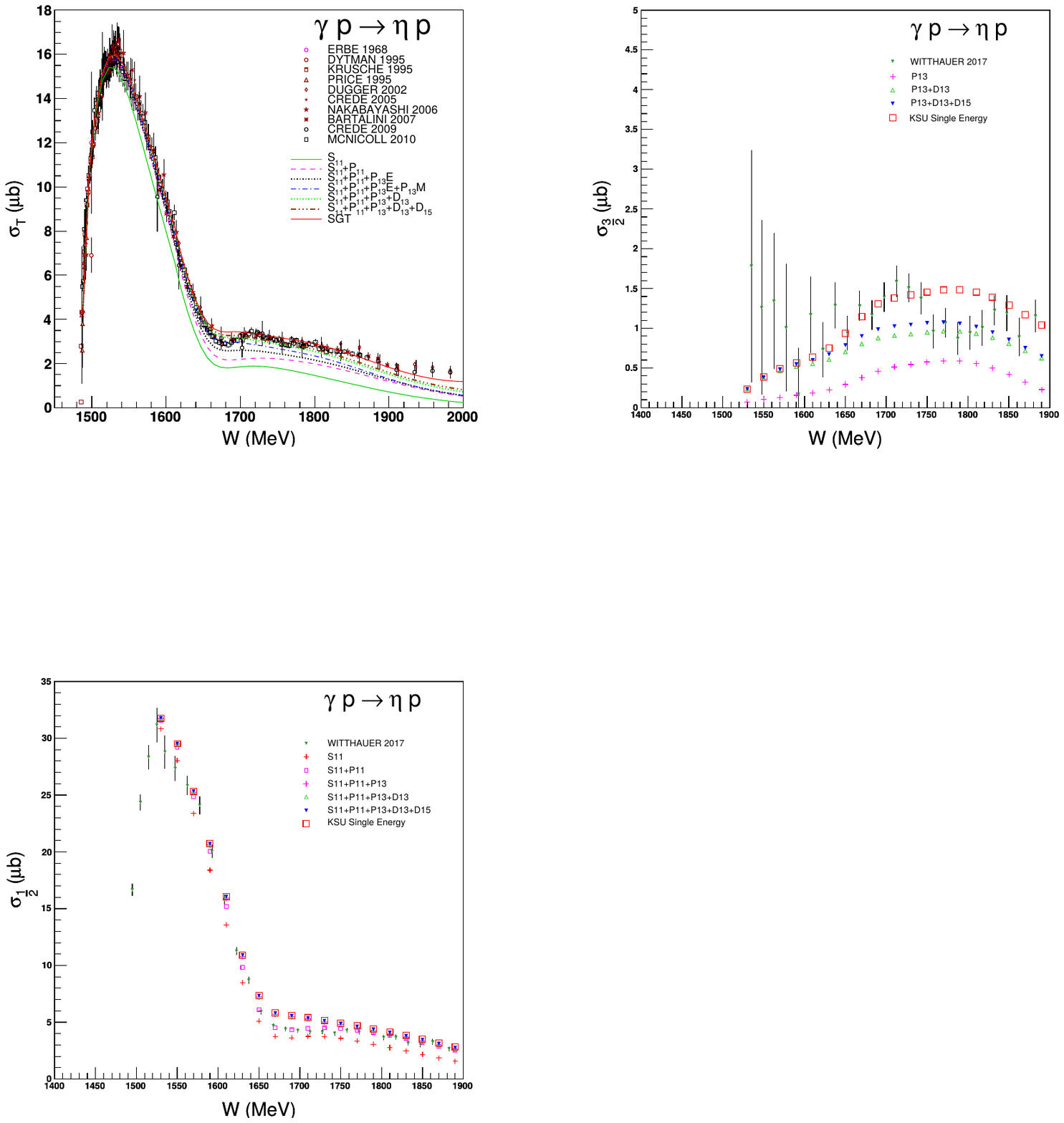}
	\caption{Helicity-$3/2$ integrated cross section for \GPEP. The data points are from Witthauer 2017 \cite{Witthauer17}. The plot also shows the contribution to the cross section by successively adding each partial wave.}
	\label{SGTGPEP32}
\end{figure}

The integrated cross section $(\sigma)$ was obtained by integrating the differential cross section over the full angular range. The helicity-1/2 and 3/2 cross sections were generated by integrating the helicity-dependent differential cross-section data as defined in Eqs.~\eqref{eqH12} and \eqref{eqH32}. Figures \ref{SGTGPEP}, \ref{SGTGPEP12}, and \ref{SGTGPEP32} show the full integrated cross section as well as the helicity-1/2 and 3/2 integrated cross sections. The curves generated through the integrated cross-section points were obtained by fitting the differential cross-section data and extracting the integrated cross section from the individual partial-wave amplitudes. As Figs.~\ref{SGTGPEP} and \ref{SGTGPEP12} show, the $\gamma p \rightarrow \eta p$ cross section rises sharply above threshold and is dominated by a bump associated with the $S_{11}(1535)$ resonance, which couples strongly to the $\eta N$ channel.  Smaller contributions come from couplings to the $P_{11}(1440)$ and $P_{11}(1880)$ resonances and the $P_{13}(1720)$ resonance.  Further details are discussed in Ref.~\cite{paper3}. Overall, the curves describe the data well through the entire energy region. The curve for $\sigma$ slightly overshoots the data near 1600~MeV but our fits to $d\sigma/d\Omega$ data were found to be in good agreement. The helicity-1/2 and helicity-3/2 plots also showed good agreement within experimental uncertainty and the scatter in the points.

\begin{figure*}

	\includegraphics[scale=0.99,trim={20mm} {44mm} {10mm} {22mm},clip=true]{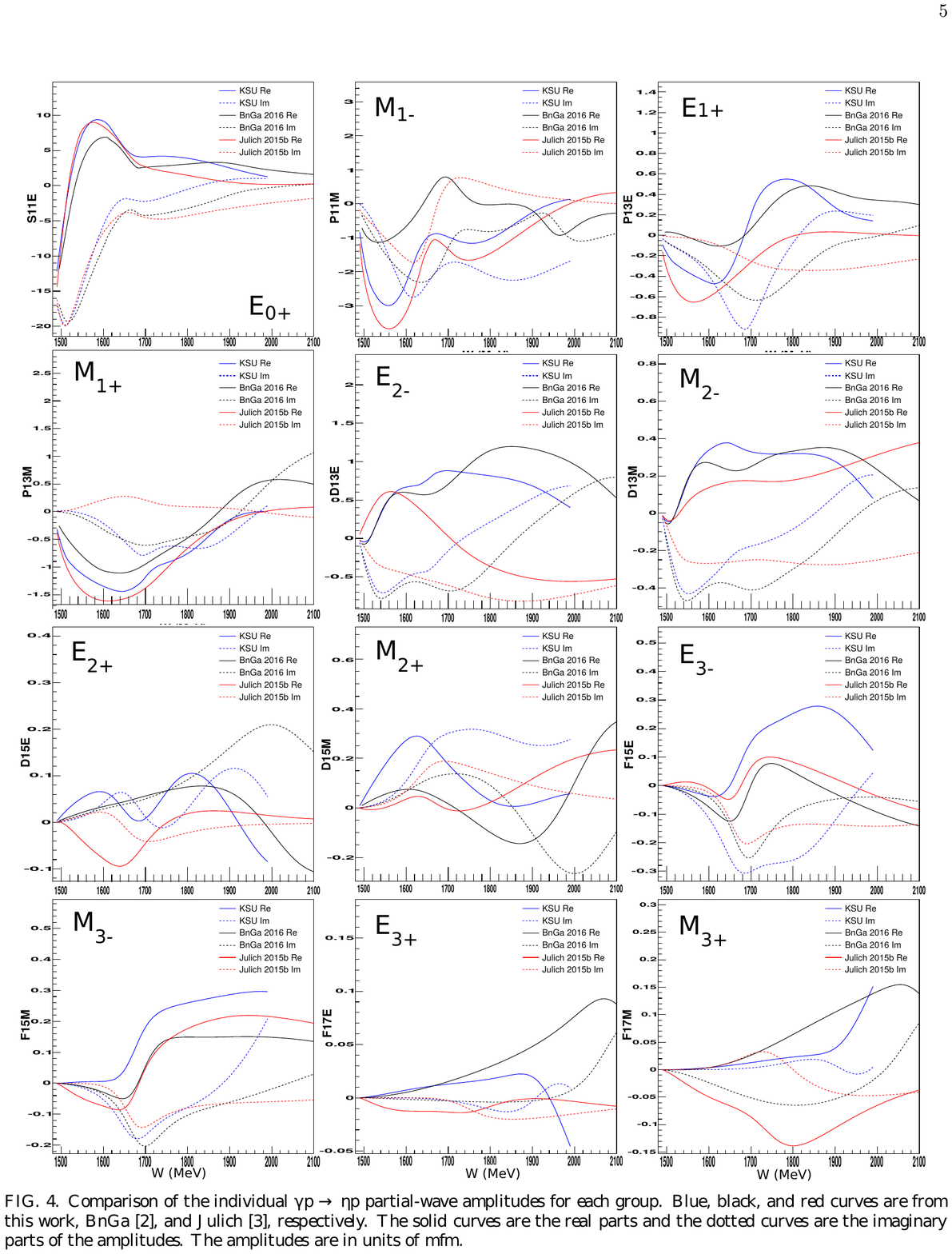}
	\caption{Comparison of the individual \GPEP \ partial-wave amplitudes for each group. Blue, black, and red curves are from this work, BnGa \cite{Sarantsev}, and Julich \cite{Julich}, respectively. The solid curves are the real parts and the dotted curves are the imaginary parts of the amplitudes. The amplitudes are in units of mfm.}	
	\label{GPEP_PWA_Compare1}
\end{figure*}

Figure \ref{GPEP_PWA_Compare1} compares the $\gamma p \rightarrow \eta p$ partial-wave amplitudes from this work with results from BnGa \cite{Sarantsev} and J\"ulich \cite{Julich}. For this reaction, the only amplitude that is in agreement between all the groups is $S_{11}$. Higher partial waves all exhibit major discrepancies with at least one of the groups. This lack of agreement indicates that additional data are needed from unmeasured double polarization observables to obtain a unique solution.

\begin{figure*}[htpb]
	\includegraphics[scale=0.99,trim={20mm} {160mm} {10mm} {24mm},clip=true]{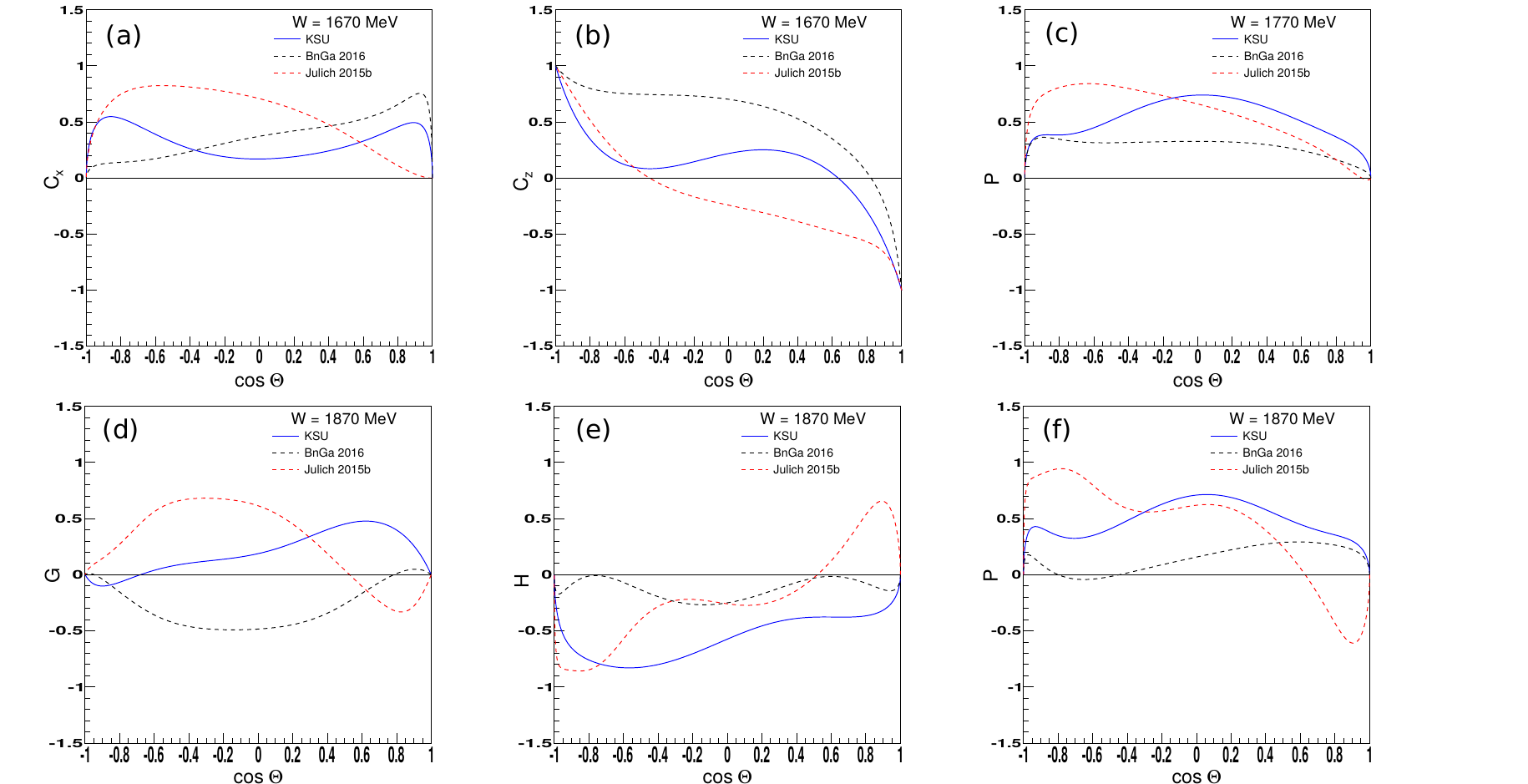}
	\caption[Predictions of $C_x$, $C_z$, $G$, $H$, and $P$ at various energies for the reaction \GPEP]{Predictions of $C_x$, $C_z$, $G$, $H$, and $P$ at various energies for the reaction \GPEP. Measurement of any of these observables at the shown energy would provide strong evidence for any needed changes in the three works results. The top two plots on the left are for $C_x$ and $C_z$, respectively, at a c.m.\ energy of 1670~MeV. The top right plot is for $P$ at 1770 MeV, and the bottom three plots are for $G$, $H$, and $P$, respectively, at 1870~MeV. The blue solid curves are from this work, the black dotted curves are from BnGa \cite{Sarantsev}, and the red dotted curves are from Julich \cite{Julich}.}	
	\label{GPEP_PREDICT_CxCz}
\end{figure*}

To obtain further progress towards the goal of a single solution for $\eta$ photoproduction off the nucleon, Fig.~\ref{GPEP_PREDICT_CxCz} shows what observables show the most difference between the three groups compared in this work. A measurement of both $C_x$ and $C_z$ would be ideal at c.m.\ energies between 1600 and 1800~MeV  while above 1800, the predictions for these two observables actually converge and a better measurement would be either $G$ or $H$. 

\subsection{$\gamma n \rightarrow \eta n$}

For the reaction \GNEN, our fits to the published observables \DSG, $E$, and $\Sigma$ are overall very good with fits to the $E$ observable are shwoing only minor local problems in a few bins. Using wide binning showed that the data varied significantly from bin-to-bin, which prevented further improvements to the fits. The $S_{11}$ amplitude dominates the reaction from threshold up to $1620$ MeV, with $P_{13}E$ and $D_{15}$ showing significant contributions in the region of the narrow structure near $1680$ MeV. Table \ref{tab:BnGavsKSU3} lists shows the $\chi^2$ contributions for this work and BnGa (2016) \cite{Sarantsev}. Again, individual and total contributions are shown. This work does a slightly better job at describing a few of the observables while BnGa (2016) does better at others. Note that preliminary data for the observables $T$ \cite{Krusche15} and $F$ \cite{Krusche15} as well as the first measurement of $E$ data published in 2017 \cite{Witthauer17} were only included in the KSU analysis.

	\begin{table}[htpb]

		\renewcommand{\arraystretch}{1.5}
		\begin{tabular}{ccc}
			\hline\hline
			Observable	& ~~~~~KSU~~~~~ & BnGa (2016) \\ \hline
			\DSG	& 6300 & 6800  \\
			$T$	    & 480 & 700 \\
			$\Sigma$& 240 & 200 \\
			$F$	    & 220 & 440 \\
			$E$	    & 250 & 150 \\
			$(d\sigma/d\Omega)_{\frac{1}{2}}$	& 310 & 260 \\
			$(d\sigma/d\Omega)_{\frac{3}{2}}$	& 210 & 140 \\
			\hline 
			Fit Total& 8100 & 8700   \\
			\hline\hline
		\end{tabular}
		\caption{\label{tab:BnGavsKSU3} $\chi^2$ contributions for $\gamma n \rightarrow \eta n$. Column~1 shows the name of the observable, columns~2 and 3 show the $\chi^2$ contributions from this work and BnGa (2016) \cite{Sarantsev}, respectively. The c.m.\ energy range was from 1490 to 1875~MeV with 5~MeV binning.}
								
	\end{table}

\begin{figure}[hptb]

	\includegraphics[scale=.99,trim={25mm} {191mm} {110mm} {19mm},clip=true]{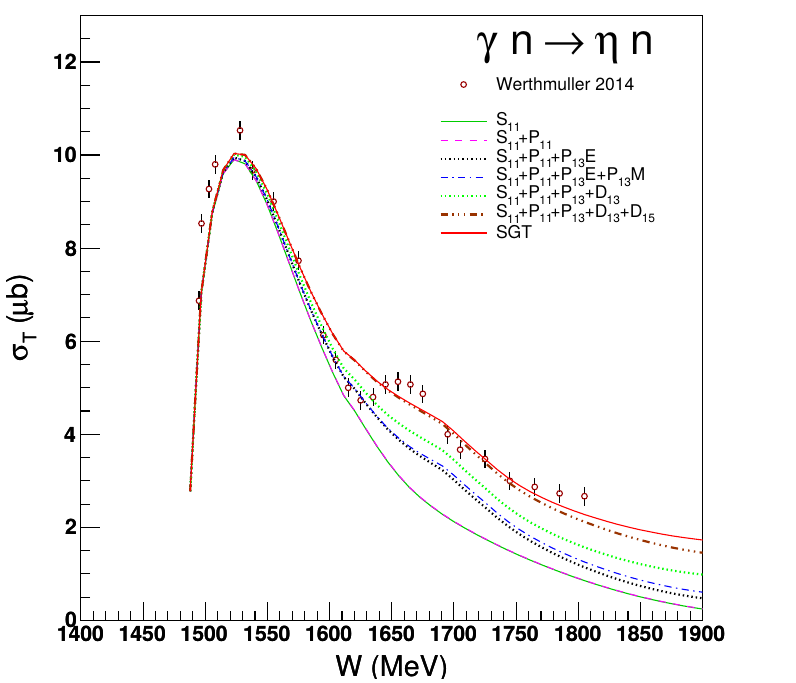}
	\caption{Integrated cross section for \GNEN. The data points are from Werthmuller 2014 \cite{Werthmuller} and the plot also shows the contribution to the cross section by successively adding each partial wave.}
	\label{SGT_GNEN}
\end{figure}

\begin{figure}[hptb]

	\includegraphics[scale=.99,trim={22mm} {102mm} {110mm} {109mm},clip=true]{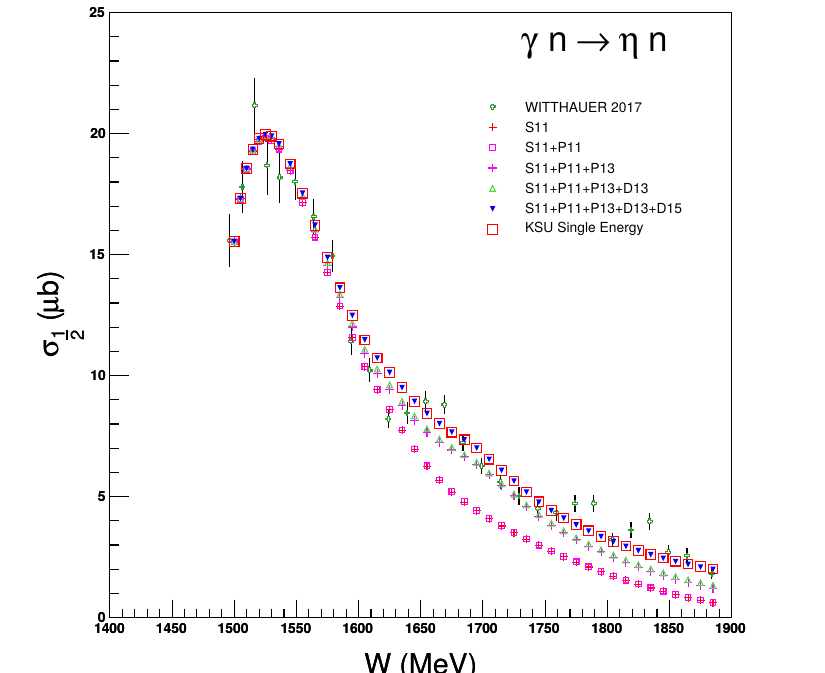}
	\caption{Helicity-$1/2$ integrated cross section for \GNEN. The data points are from Witthauer 2017 \cite{Witthauer17} and the plot also shows the contribution to the cross section by successively adding each partial wave.}
	\label{SGT_GNEN12}
\end{figure}

\begin{figure}[hptb]
	\includegraphics[scale=.99,trim={113mm} {191mm} {\trimCC} {\trimDD},clip=true]{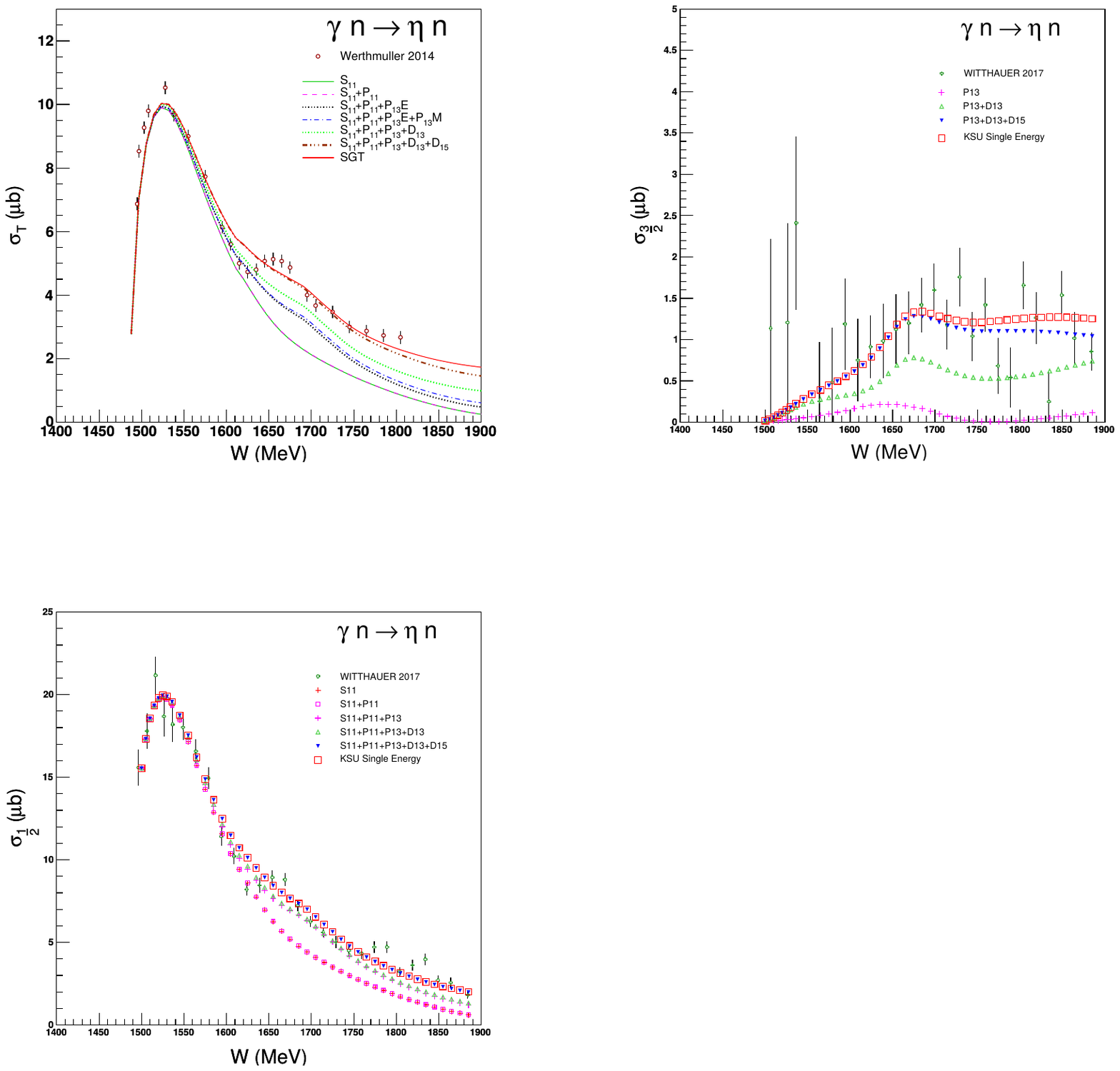}
	\caption{Helicity-$3/2$ integrated cross section for \GNEN. The data points are from Witthauer 2017 \cite{Witthauer17} and the plot also shows the contribution to the cross section by successively adding each partial wave.}
	\label{SGT_GNEN32}
\end{figure}

Figures \ref{SGT_GNEN}, \ref{SGT_GNEN12}, and \ref{SGT_GNEN32} show integrated cross-section results.  The dominant structure in Figs.~\ref{SGT_GNEN} and \ref{SGT_GNEN12} is the bump associated with the $S_{11}(1535)$ resonance, which couples strongly to the $\eta N$ channel.  Both of these figures also reveal what appears to be a narrow structure less than 100~MeV wide near 1680~MeV. Much has been written about this structure. Some researchers have concluded that the bump must be from either a $P_{11}$ resonance or due to an interference effect between two $S_{11}$ resonances \cite{Kuznetsov, Anisovich1685} because the structure only appears in the helicity-$1/2$ data (see Figs.~\ref{SGT_GNEN12} and \ref{SGT_GNEN32}). The argument has been made that the helicity-$1/2$ cross section contains contributions from $S_{11}$ and $P_{11}$, while the helicity-$3/2$ cross section does not, so the bump must be due to these two amplitudes. 

The present work provides an alternative interpretation of the bump as a complicated structure generated by a number of resonances, specifically $D_{13}(1700)$, $D_{15}(1675)$, and the tail of the $S_{11}(1535)$. As mentioned above, this would generate a bump in the helicity-3/2 cross section. Our predictions of the helicity cross sections (Figs.~\ref{SGT_GNEN12} and \ref{SGT_GNEN32}) show that the data allow and even hint at a small bump within the size of the error bars and scatter of the points. While the fits to the integrated cross section seem to overshoot the data at c.m.\ energies near 1650~MeV, the energy resolution in the region around the bump is 30~MeV (and wider at higher energies), despite the cross-section points being roughly 10~MeV apart \cite{Werthmuller}.  Further details about the resonance content (masses, widths, branching ratios, pole positions, \textit{etc.}) can be found in Ref.~\cite{paper3}.

\begin{figure*}[hbpt]

	\includegraphics[scale=0.99,trim={20mm} {44mm} {\trimCC} {\trimDD},clip=true]{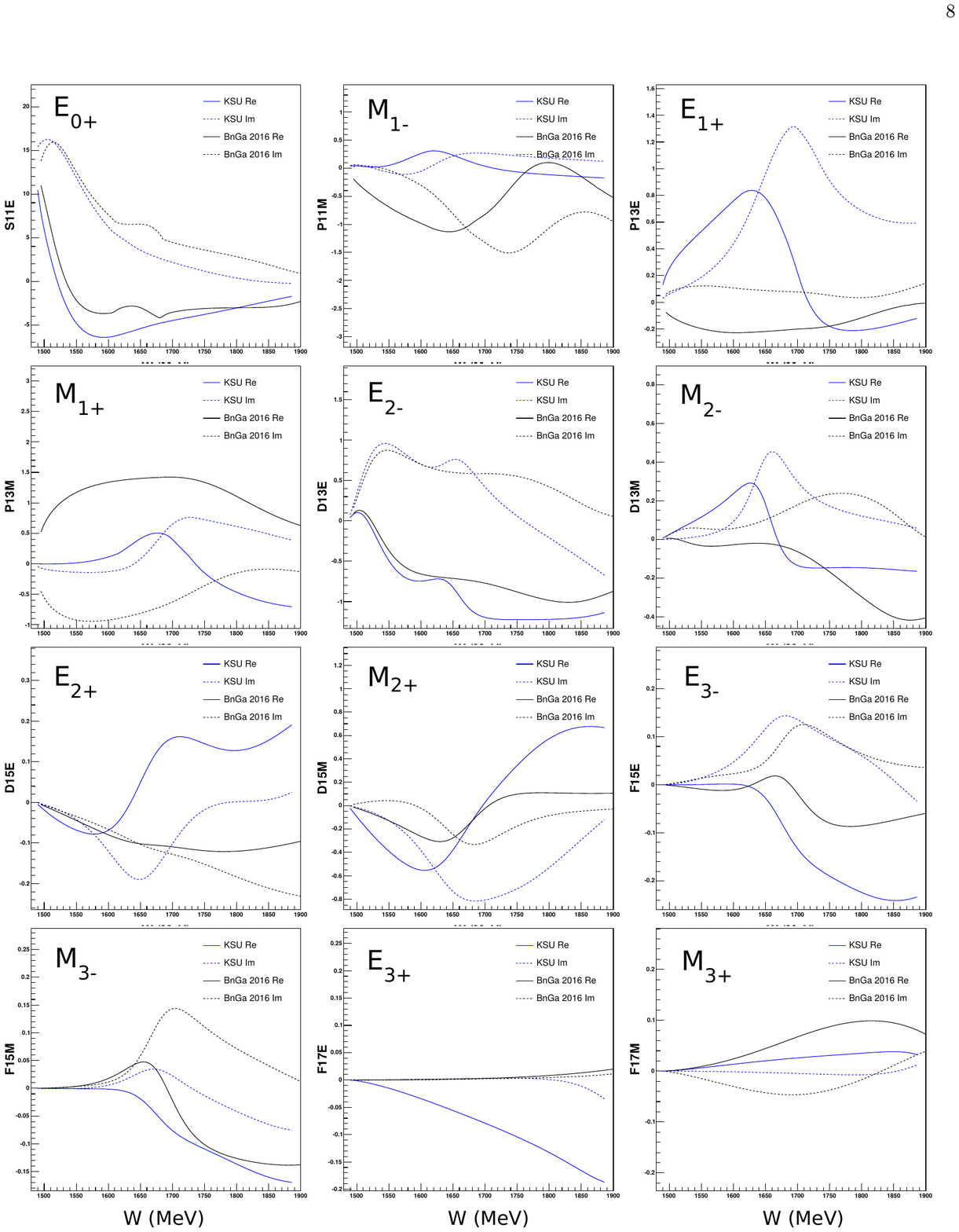}
	\caption{Comparison of the individual \GNEN \ partial-wave amplitudes for each group. The blue and black curves are from this work and BnGa \cite{Sarantsev}, respectively. The solid curve is the real part and the dotted curve the imaginary part of the amplitude. The amplitudes are in units of mfm.}	
	\label{GNEN_PWA_Compare1}
\end{figure*}

Plots comparing the \GNEN \ partial-wave amplitudes determined in this work with BnGa results \cite{Sarantsev} are shown in Figure \ref{GNEN_PWA_Compare1}. The $S_{11}$ amplitude is similar except in the c.m.\ energy region near 1680~MeV where the  BnGa group explains the bump as an $S_{11}$ interference and what looks like a cusp effect, possibly from the opening of the $K \Sigma$ channel near 1680~MeV. The only other amplitude that is similar between the two groups is the $D_{13}E$ amplitude.

In our fits, there appears to be more structure than that found in the BnGa solution. Resonance peaks are clearly seen in multiple amplitudes with the imaginary part of the amplitude forming a peak when the corresponding real part approaches zero, as expected for a Breit-Wigner resonance.

At this point, additional measurements of any of the 16 observables would be useful to constrain the fits further and confirm previous measurements. As such, no predictions are shown at this point.

\clearpage
\section{Summary and Conclusions}
Results from a partial-wave analysis of available data for $\gamma N \rightarrow \eta N$ were presented. $S_{11}$, $P_{11}$, $P_{13}$, and $F_{15}$ amplitudes were found to be important for the the reaction $\gamma p \rightarrow \eta p$ in the energy range from threshold to 2000~MeV. $S_{11}$, $P_{13}$, $D_{13}$, and $D_{15}$ were important for \GNEN. This is consistent with the Moorhouse selection rule \cite{Moorhouse}, which predicts that the $D_{15}$(1675) resonance may couple to $\gamma n$ but not to $\gamma p$. 

Also, despite the wealth of new data, measurements of additional double polarization measurements are still needed to obtain agreement between the different partial-wave analyses.

The $\gamma p \to \eta p$  and $\gamma n \rightarrow \eta n$ amplitudes from this work have been included in an updated multichannel energy-dependent partial-wave analysis \cite{paper3} that also incorporates our single-energy amplitudes for $\gamma p \rightarrow K^+ \Lambda$ \cite{paper2}.  In Ref.~\cite{paper3}, we present and discuss the resonance parameters obtained from a fit of single-energy amplitudes for these reactions combined with corresponding amplitudes for $\gamma N \rightarrow \pi N$, $\pi N \rightarrow \pi N$, $\pi N \rightarrow \pi \pi N$, $\pi N \rightarrow K \Lambda$, and $\pi N \rightarrow \eta N$.  Reference~\cite{paper3} also includes Argand diagrams that compare the results of our single-energy fits with our final energy-dependent partial-wave amplitudes.

\begin{acknowledgments}

The authors would like to thank Professor Igor Strakovsky for supplying much of the database. The authors also thank Professor Bernd Krusche for providing us with preliminary results for some of the data we fitted.  This work was supported in part by the U.S. Department of Energy, Office of Science, Office of Nuclear Physics Research Division, under Awards No.\ DE-FG02-01ER41194 and DE-SC0014323, and by the Department of Physics at Kent State University. 

\end{acknowledgments}




\appendix
\section{Final Fits to Experimental Data}
Figures \ref{GPEP_Fig1} - \ref{GPEP_Fig30} show fits to {\GPEP} and {\GNEN} data for the observables $d\sigma/d\Omega$, $\Sigma$, $T$, $F$, and $E$. 
The partial-wave amplitudes used to generate the curves are available in the form of data files \cite{Datafile}.
Also shown are the fits from BnGa 2016 \cite{Sarantsev} and J\"ulich 2015b \cite{Julich}. 

Data sources shown in the plots of \GPEP \ are: HEUSCH 1966 \cite{Heusch66},  DELCOURT 1969 \cite{Delcourt69},   CHRIST 1973 \cite{Christ73}, BOOTH 1974 \cite{Booth74},  VARTAPETYAN 1980 \cite{Vartapetyan80}, HOMMA 1988 \cite{Homma88}, DYTMAN 1995 \cite{Dytman95},  KRUSCHE 1995 \cite{Krusche95}, PRICE 1995 \cite{Price95},  AJAKA 1998 \cite{Ajaka98}, KOUZNETSOV 1998 \cite{Kouznetsov98}, DUGGER 2002 \cite{Dugger02}, AHRENS 2003 \cite{Ahrens03}, CREDE 2005 \cite{Crede05}, NAKABAYASHI 2006 \cite{Nakbayashi06}, BARTALINI 2007 \cite{Bartalini07}, ELSNER 2007 \cite{Elsner07}, SUMIHAMA 2009 \cite{Sumihama09}, WILLIAMS 2009 \cite{Williams09}, CREDE 2009 \cite{Crede09}, MCNICOLL 2010 \cite{Mcnicoll10},  AKONDI 2014 \cite{Akondi14}, SENDEROVICH 2014 \cite{Senderovich14}, KASHEVAROV 2016 \cite{Kashevarov17}, and WITTHAUER 2017 \cite{Witthauer17}.  


Data sources shown in the plots of \GNEN \ are FANTINI 2008, \cite{Fantini}, WERTHMULLER 2014 \cite{Werthmuller}, and WITTHAUER 2017 \cite{Witthauer17}.
\newcommand{\scalefac}{0.96}
\onecolumngrid
\noindent
\begin{figure}
	\includegraphics[scale=\scalefac,trim={18mm} {30mm} {\trimC} {13mm},clip=true]{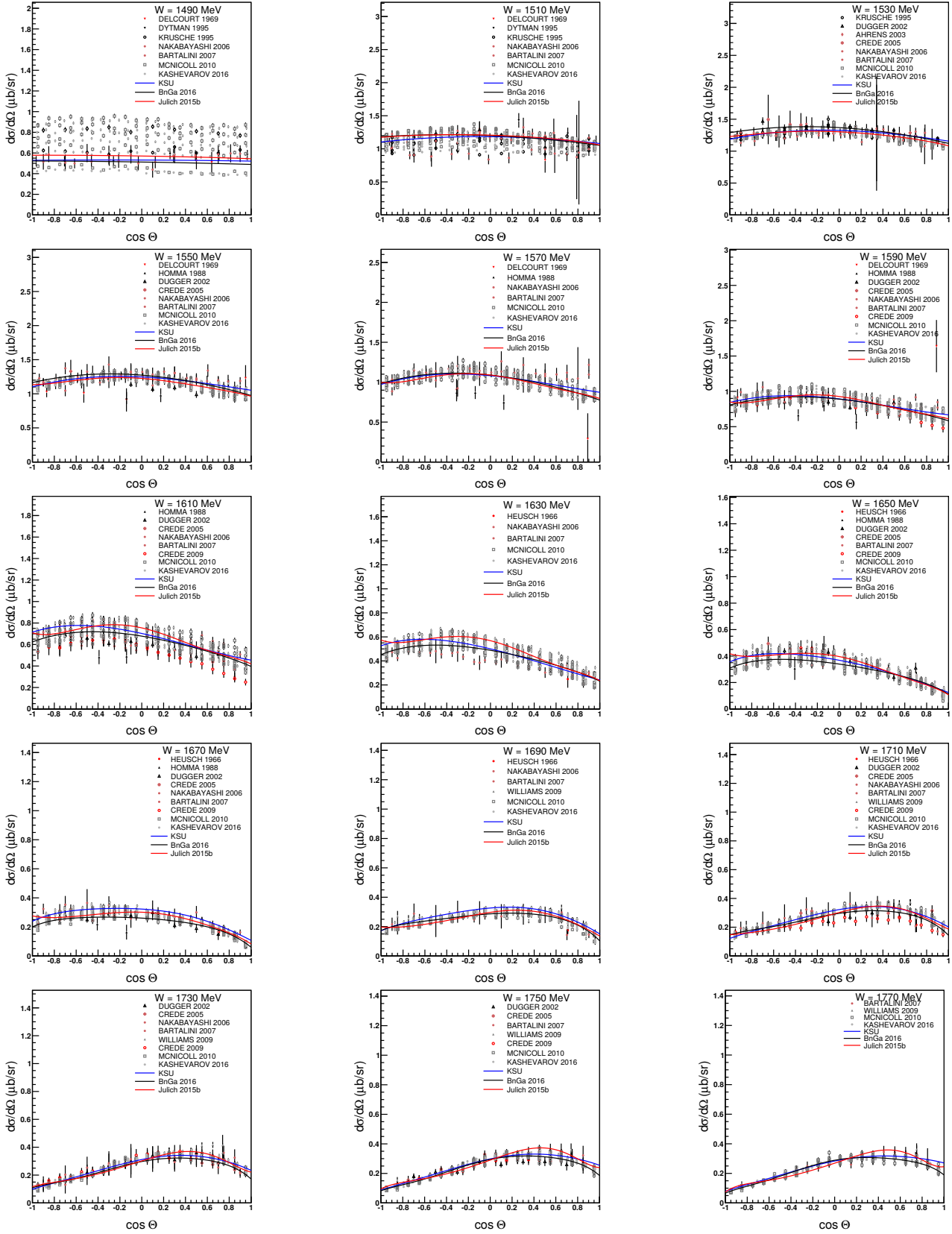}

\caption{\label{GPEP_Fig1}Fits to $d\sigma/d\Omega$ data for $\gamma p \rightarrow \eta p$ at $W$ = 1490 to 1770~MeV. See text for references. }
\end{figure}

\begin{figure}

	\includegraphics[scale=\scalefac,trim={18mm} {48mm} {\trimC} {13mm},clip=true]{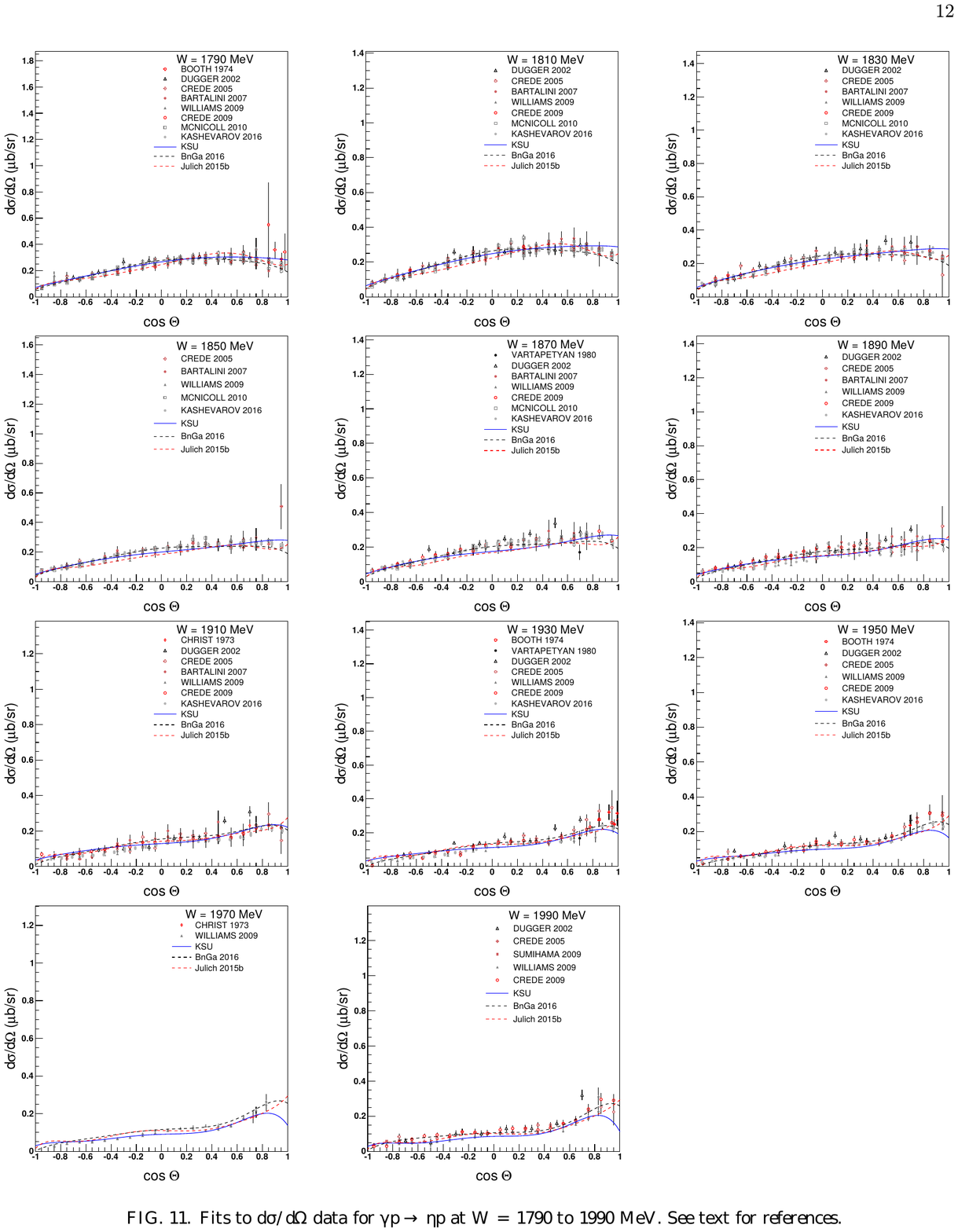}
	\caption{Fits to $d\sigma/d\Omega$ data for  $\gamma p \rightarrow \eta p$ at $W$ = 1790 to 1990~MeV. See text for references. }
\end{figure}

	\renewcommand{\ObsName}{S}
	\renewcommand{\printObs}{$\Sigma$}
\begin{figure}
	\label{GPEP_Fig5}
	\includegraphics[scale=\scalefac,trim={18mm} {48mm} {\trimC} {13mm},clip=true]{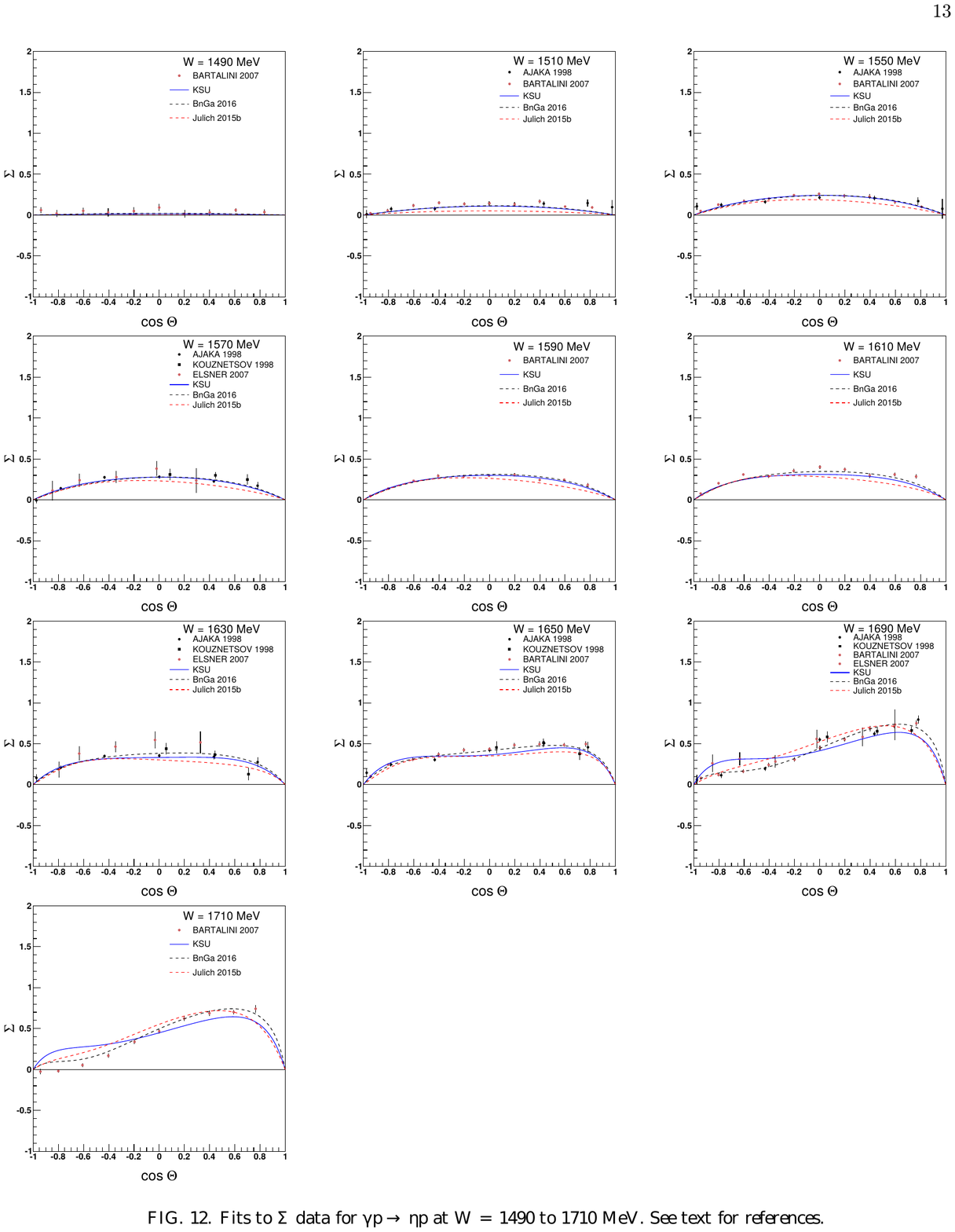}
	\caption{Fits to $\Sigma$ data for $\gamma p \rightarrow \eta p$ at $W$ = 1490 to 1710~MeV.  See text for references.}
\end{figure}
\noindent
\begin{figure}
	\label{GPEP_Fig8}
	\includegraphics[scale=\scalefac,trim={18mm} {100mm} {\trimC} {13mm},clip=true]{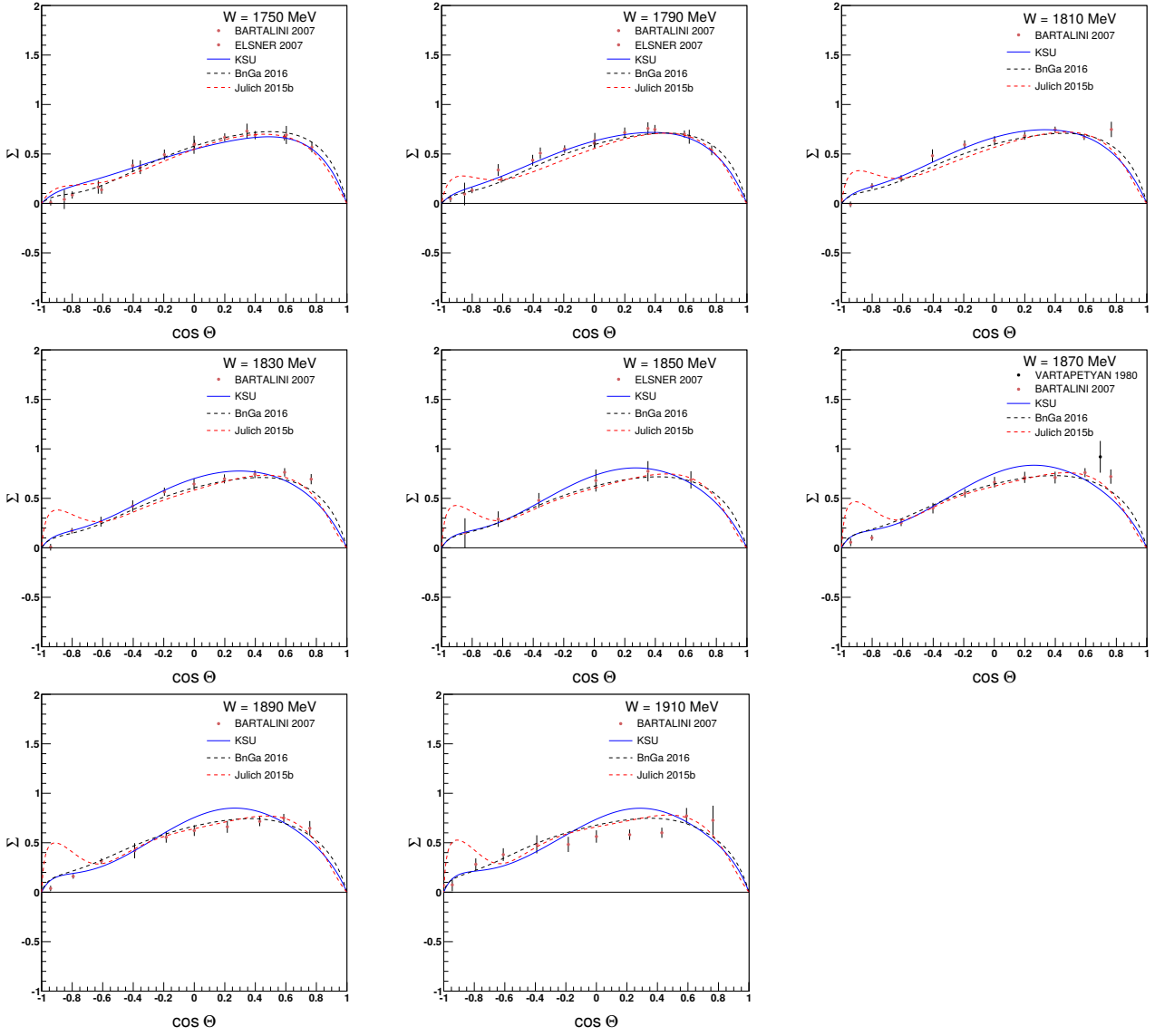}	
	\caption{Fits to $\Sigma$ data for  $\gamma p \rightarrow \eta p$ at $W$ = 1750 to 1910~MeV. See text for references. }
\end{figure}%

	\renewcommand{\ObsName}{T}
	\renewcommand{\printObs}{$T$}
\noindent
\begin{figure}
	\label{GPEP_Fig9}

	\includegraphics[scale=\scalefac,trim={18mm} {48mm} {\trimC} {13mm},clip=true]{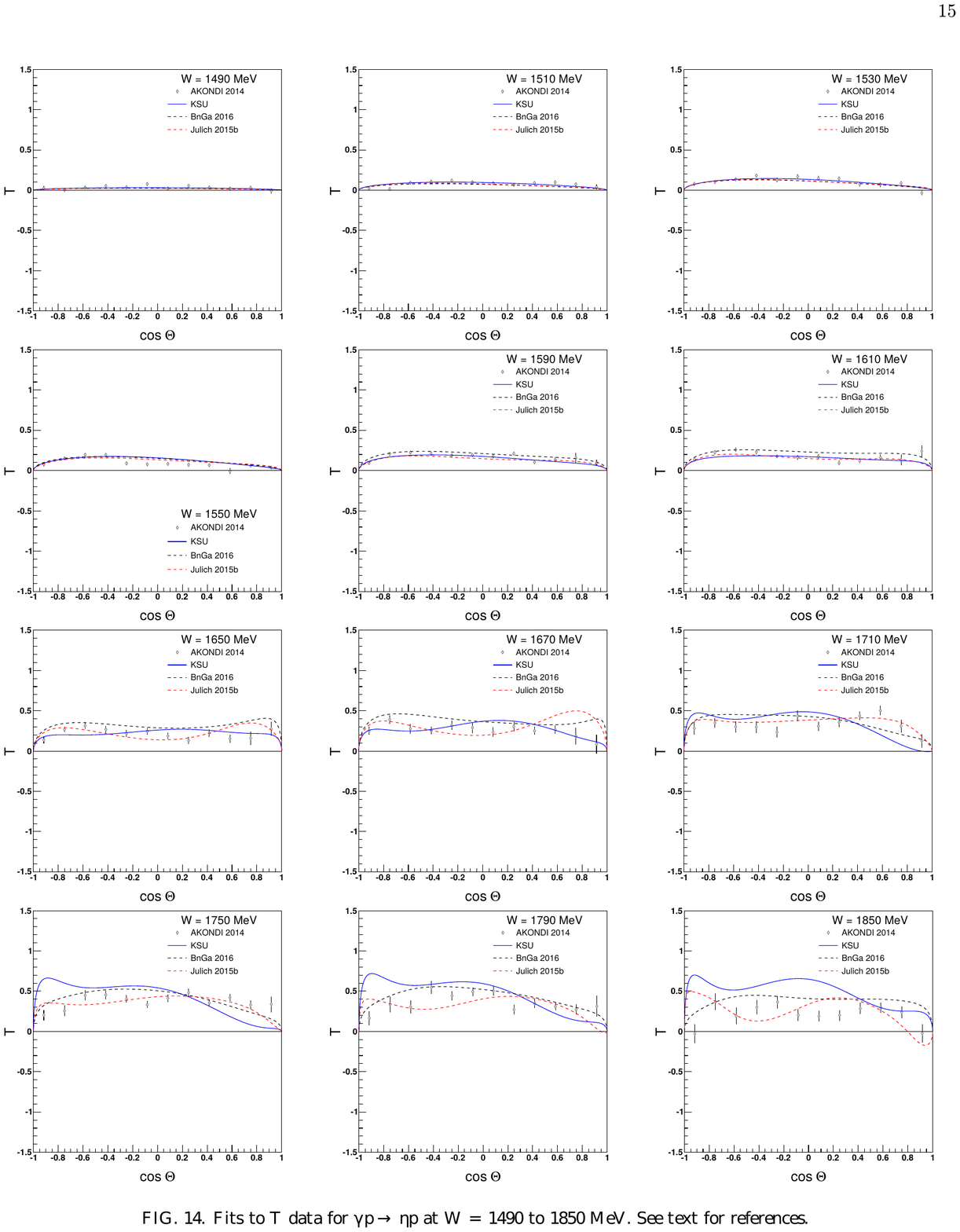}
	\caption{Fits to $T$ data for $\gamma p \rightarrow \eta p$  at $W$ = 1490 to 1850~MeV. See text for references. }
\end{figure}

	\renewcommand{\ObsName}{F}
	\renewcommand{\printObs}{$F$}
	\noindent
	\begin{figure}
		\label{GPEP_Fig10}
		
	\includegraphics[scale=\scalefac,trim={18mm} {48mm} {\trimC} {13mm},clip=true]{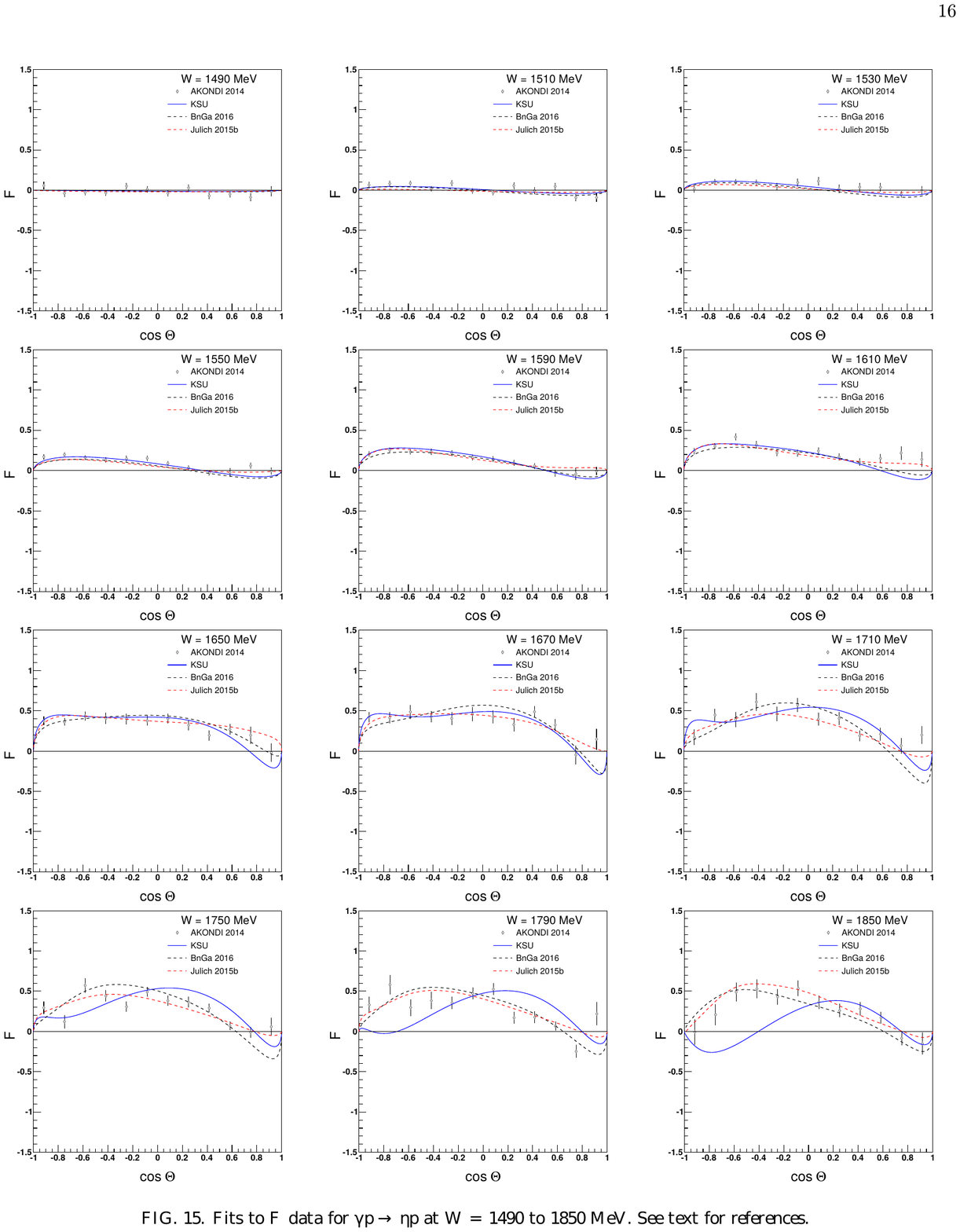}
		\caption{Fits to $F$ data for $\gamma p \rightarrow \eta p$  at $W$ = 1490 to 1850~MeV. See text for references. }
	\end{figure}

\renewcommand{\ObsName}{E}
\renewcommand{\printObs}{$E$}
\noindent
\begin{figure}
	\label{GPEP_Fig13}
	\includegraphics[scale=\scalefac,trim={18mm} {50mm} {\trimC} {13mm},clip=true]{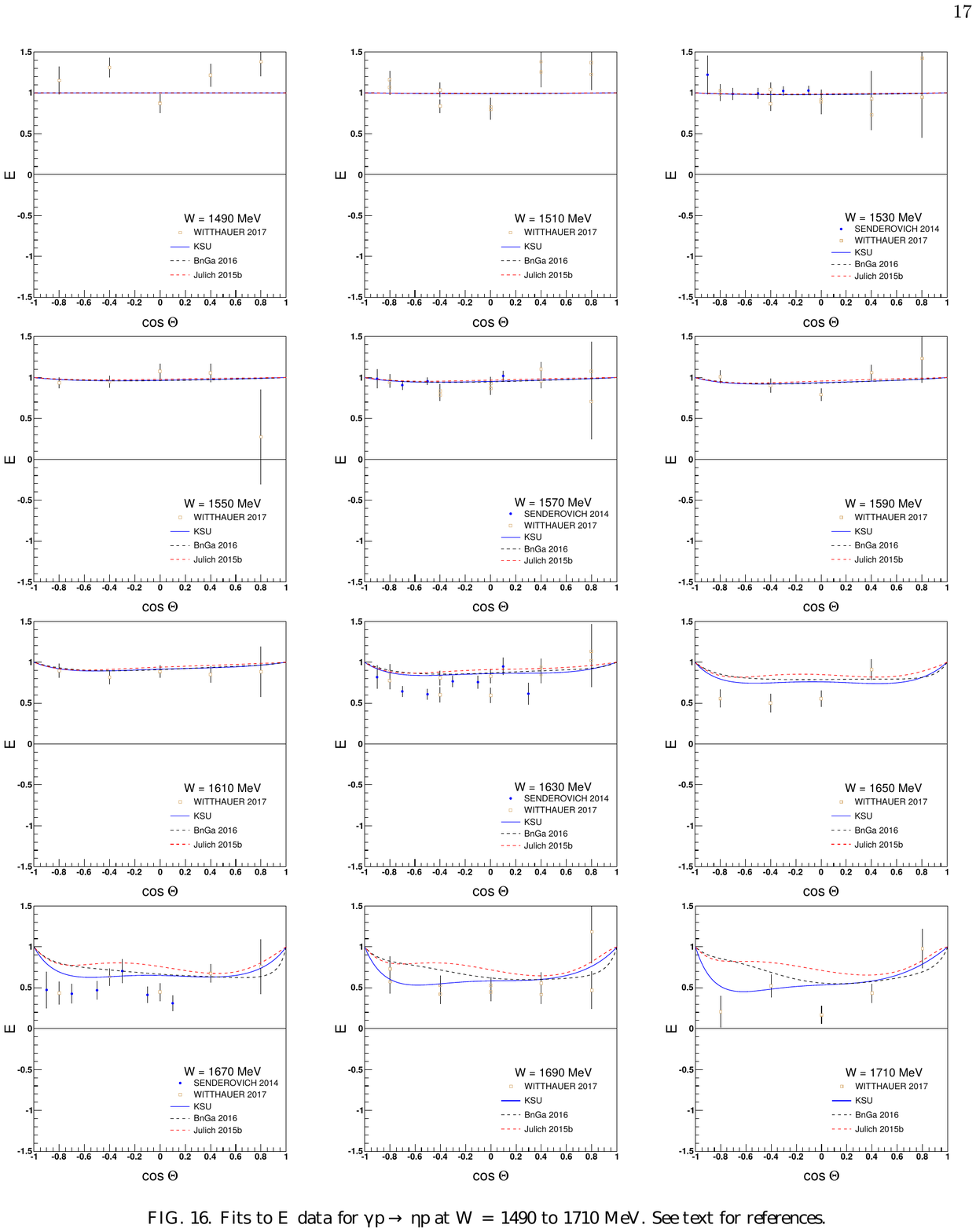}
	\caption{Fits to $E$ data for  $\gamma p \rightarrow \eta p$  at $W$ = 1490 to 1710~MeV. See text for references.}
\end{figure}
\clearpage

\noindent
\begin{figure}
	\label{GPEP_Fig27}
	\includegraphics[scale=\scalefac,trim={18mm} {50mm} {\trimC} {13mm},clip=true]{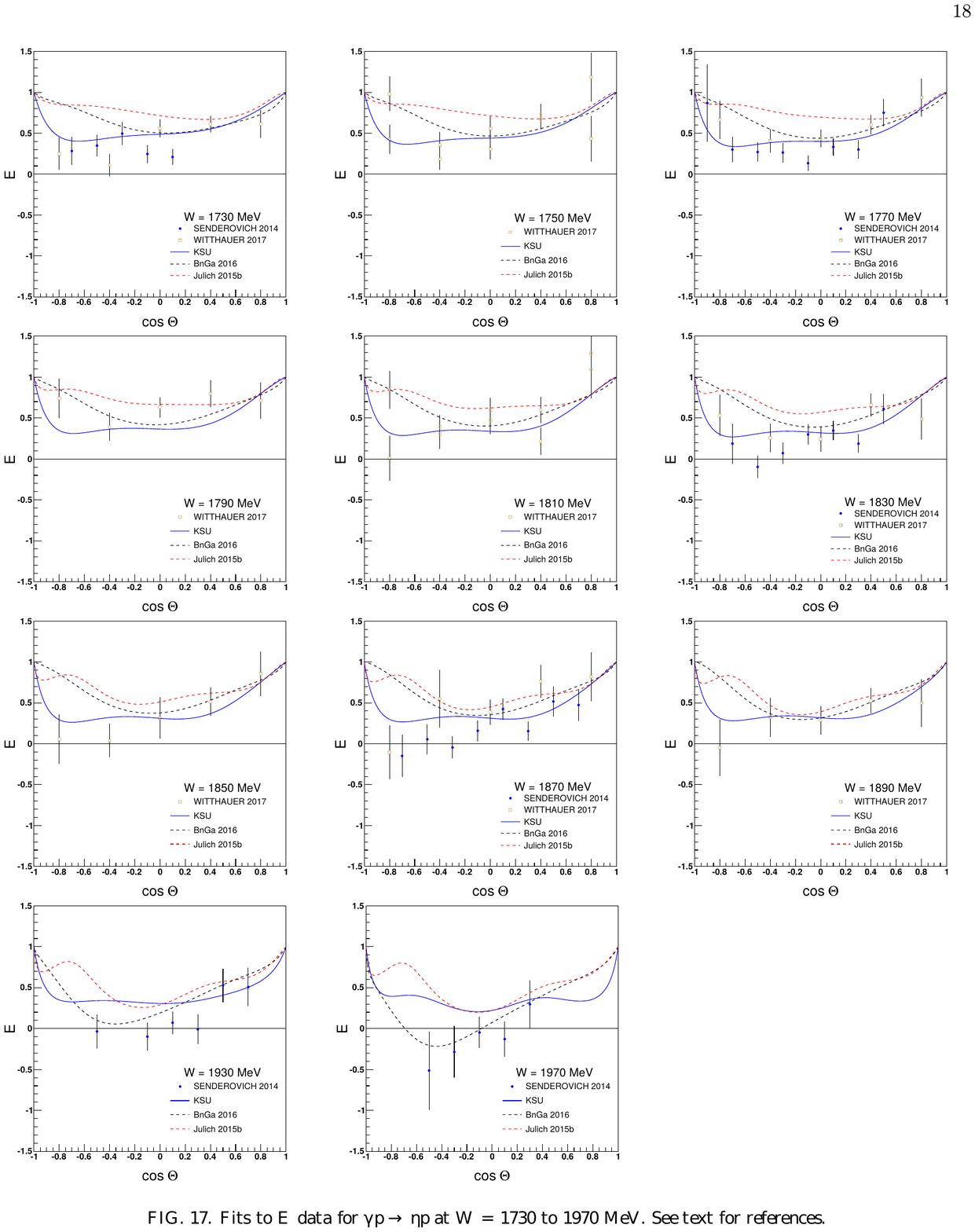}		
	\caption{Fits to $E$ data for $\gamma p \rightarrow \eta p$  at $W$ = 1730 to 1970~MeV. See text for references.}
\end{figure}
\renewcommand{\pathGraphing}{\pathGNEN}
\renewcommand{\printReac}{\GNEN}
\renewcommand{\ObsName}{DSG}
\renewcommand{\printObs}{\DSGT}
\noindent
\begin{figure}
	\includegraphics[scale=\scalefac,trim={18mm} {22mm} {\trimC} {13mm},clip=true]{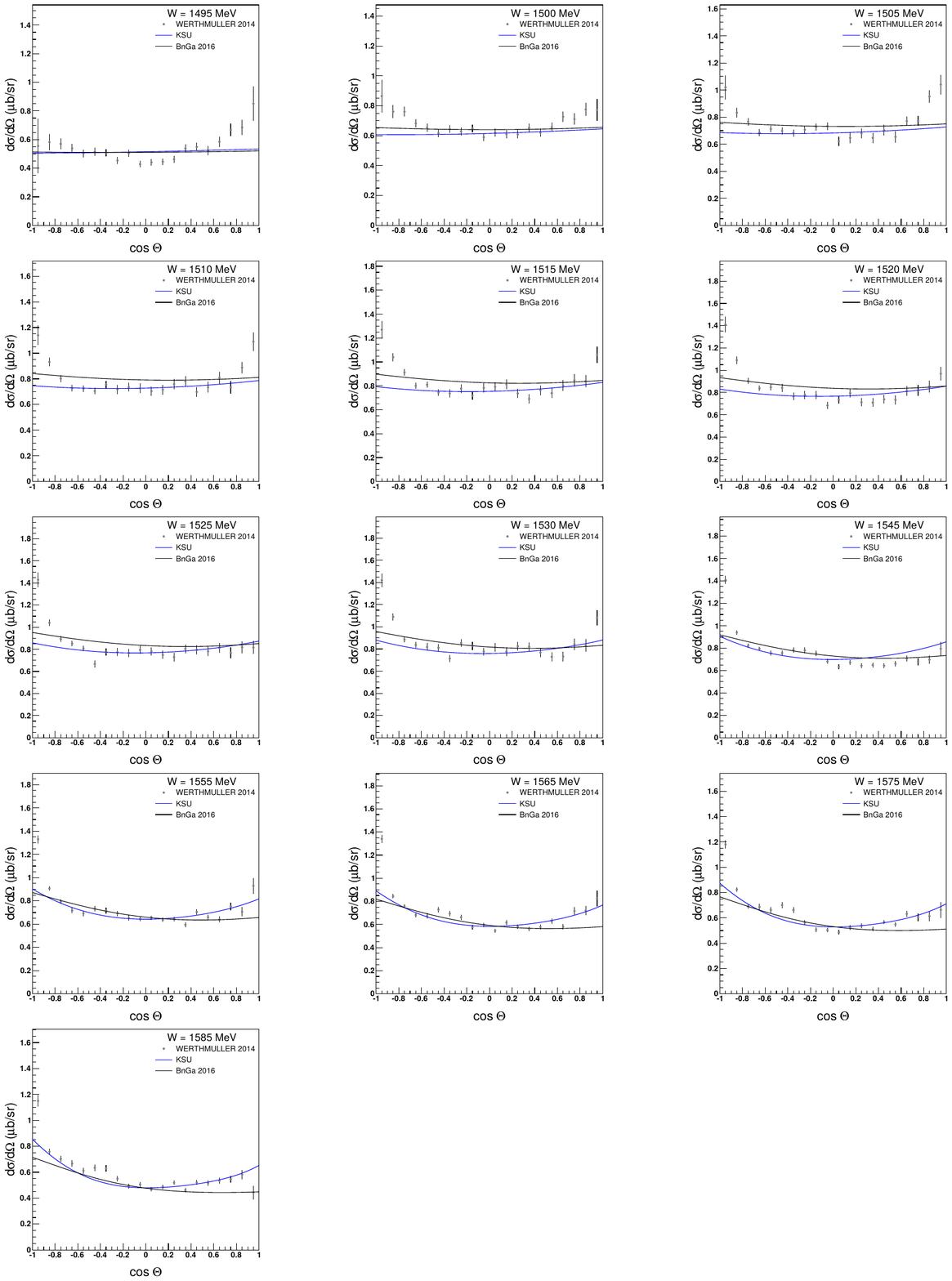}
	\caption{Fits to $d\sigma/d\Omega$ data for $\gamma n \rightarrow \eta n$ at $W$ = 1495 to 1585~MeV. See text for references. }
	\label{GNENFig1}
\end{figure}%

\noindent

\begin{figure}
	\includegraphics[scale=\scalefac,trim={18mm} {22mm} {\trimC} {13mm},clip=true]{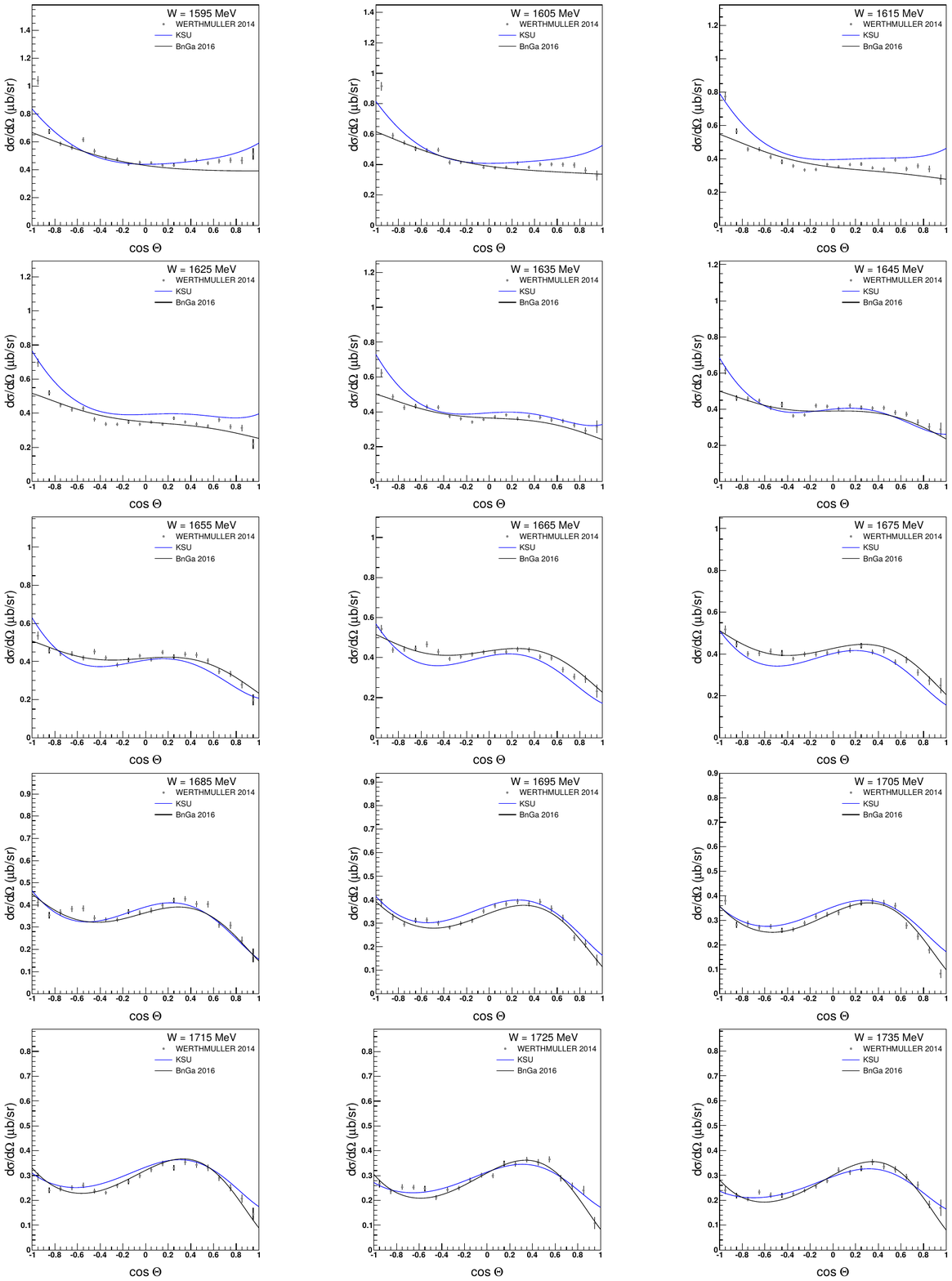}
	\caption{Fits to $d\sigma/d\Omega$ data for  $\gamma n \rightarrow \eta n$ at $W$ = 1595 to 1735~MeV. See text for references.}
\end{figure}

\noindent
\begin{figure}
	\includegraphics[scale=\scalefac,trim={18mm} {22mm} {\trimC} {13mm},clip=true]{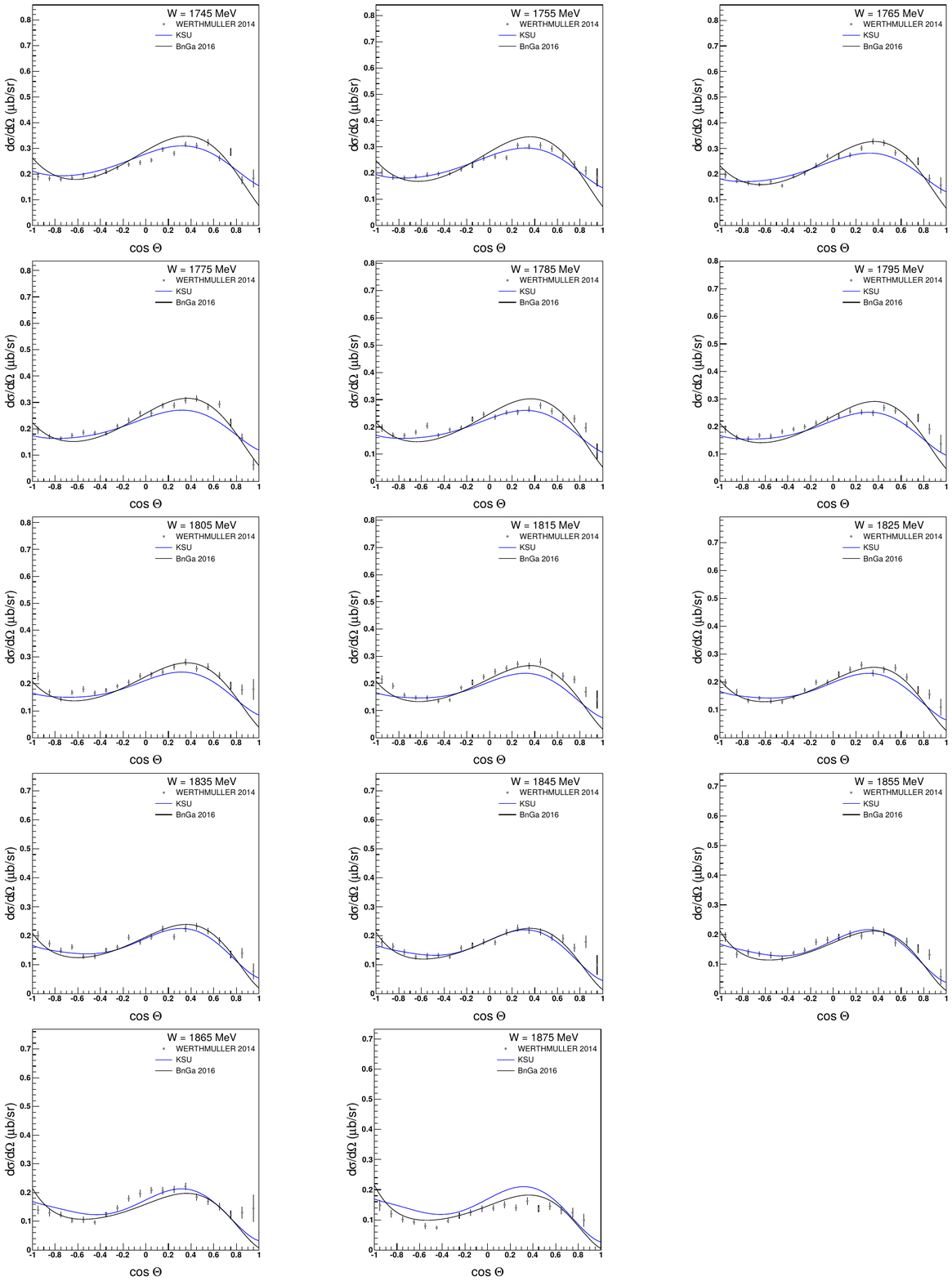}	
	\caption{Fits to $d\sigma/d\Omega$ data for $\gamma n \rightarrow \eta n$ at $W$ = 1745 to 1875~MeV. See text for references.}
\end{figure}

	\renewcommand{\ObsName}{S}
	\renewcommand{\printObs}{$\Sigma$}
\noindent
\begin{figure}
	\includegraphics[scale=\scalefac,trim={18mm} {70mm} {\trimC} {13mm},clip=true]{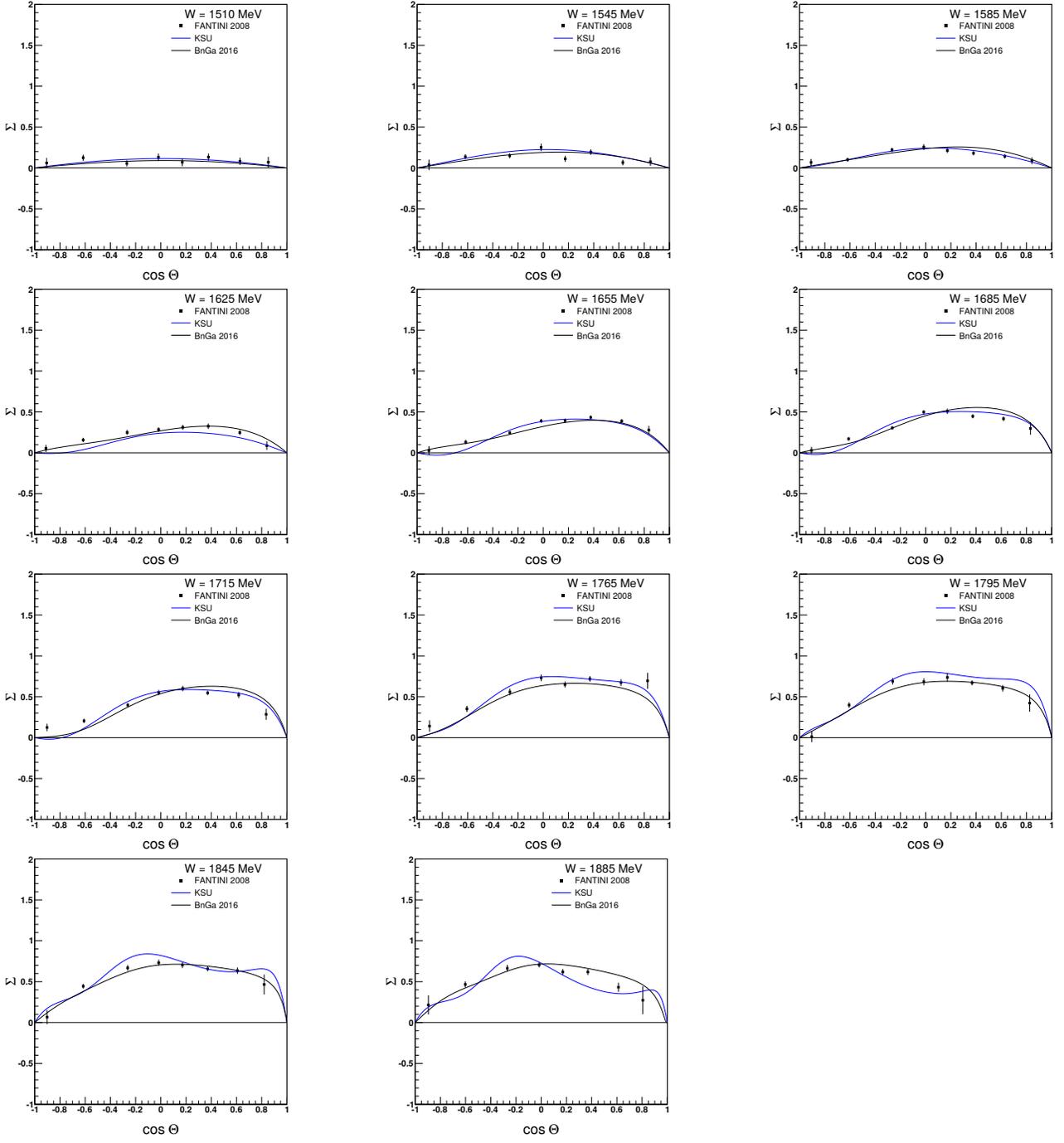}

	\caption{Fits to $\Sigma$ data for $\gamma n \rightarrow \eta n$ at $W$ = 1510 to 1885~MeV. See text for references.}
\end{figure}


\renewcommand{\ObsName}{E}
\renewcommand{\printObs}{$E$}
\noindent
\begin{figure}
	\includegraphics[scale=\scalefac,trim={18mm} {22mm} {\trimC} {13mm},clip=true]{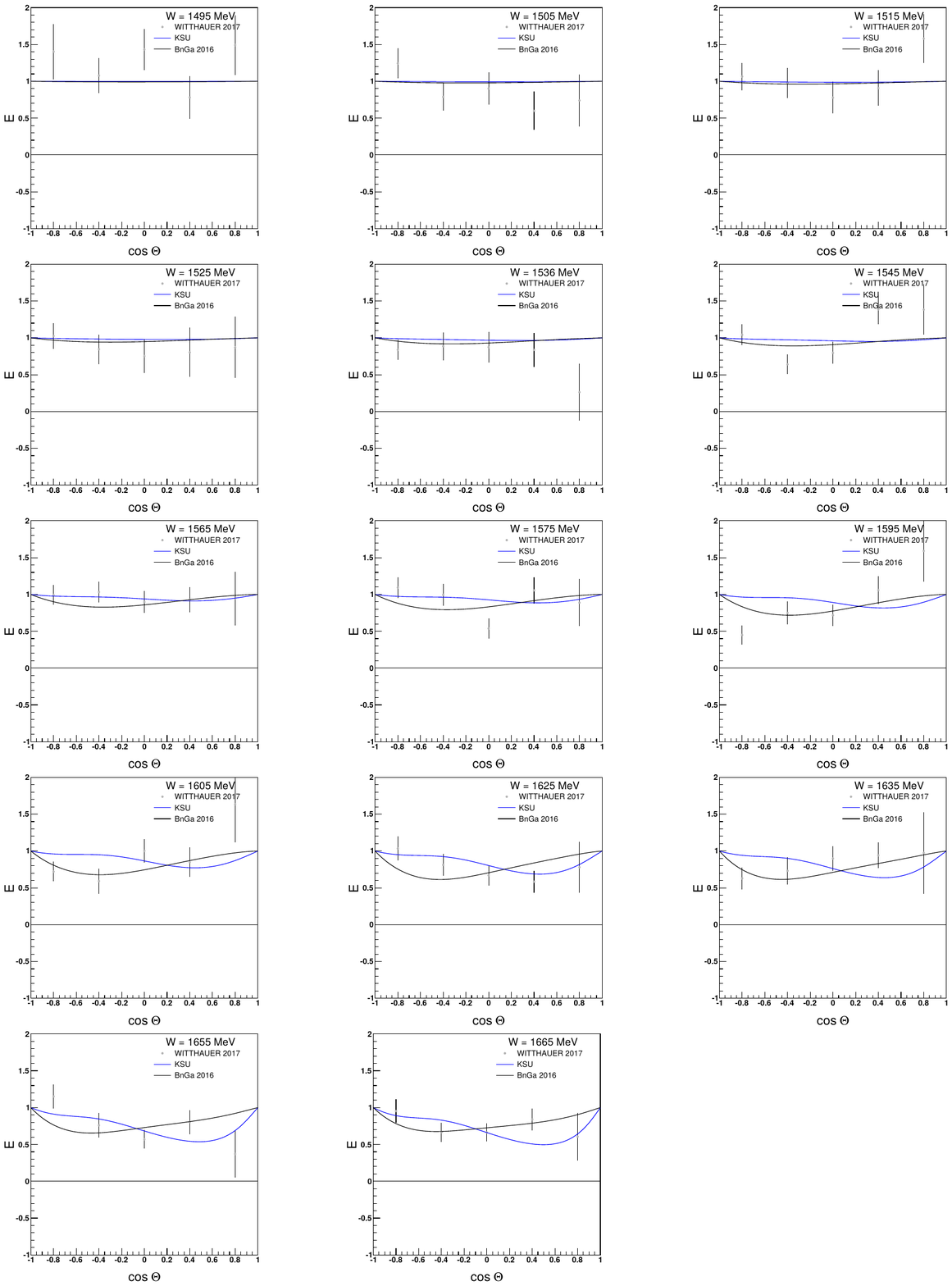}	
	\caption{Fits to $E$ data for $\gamma n \rightarrow \eta n$ at $W$ = 1495 to 1665~MeV. See text for references.}
\end{figure}

\noindent
\begin{figure}
	\includegraphics[scale=\scalefac,trim={18mm} {22mm} {\trimC} {13mm},clip=true]{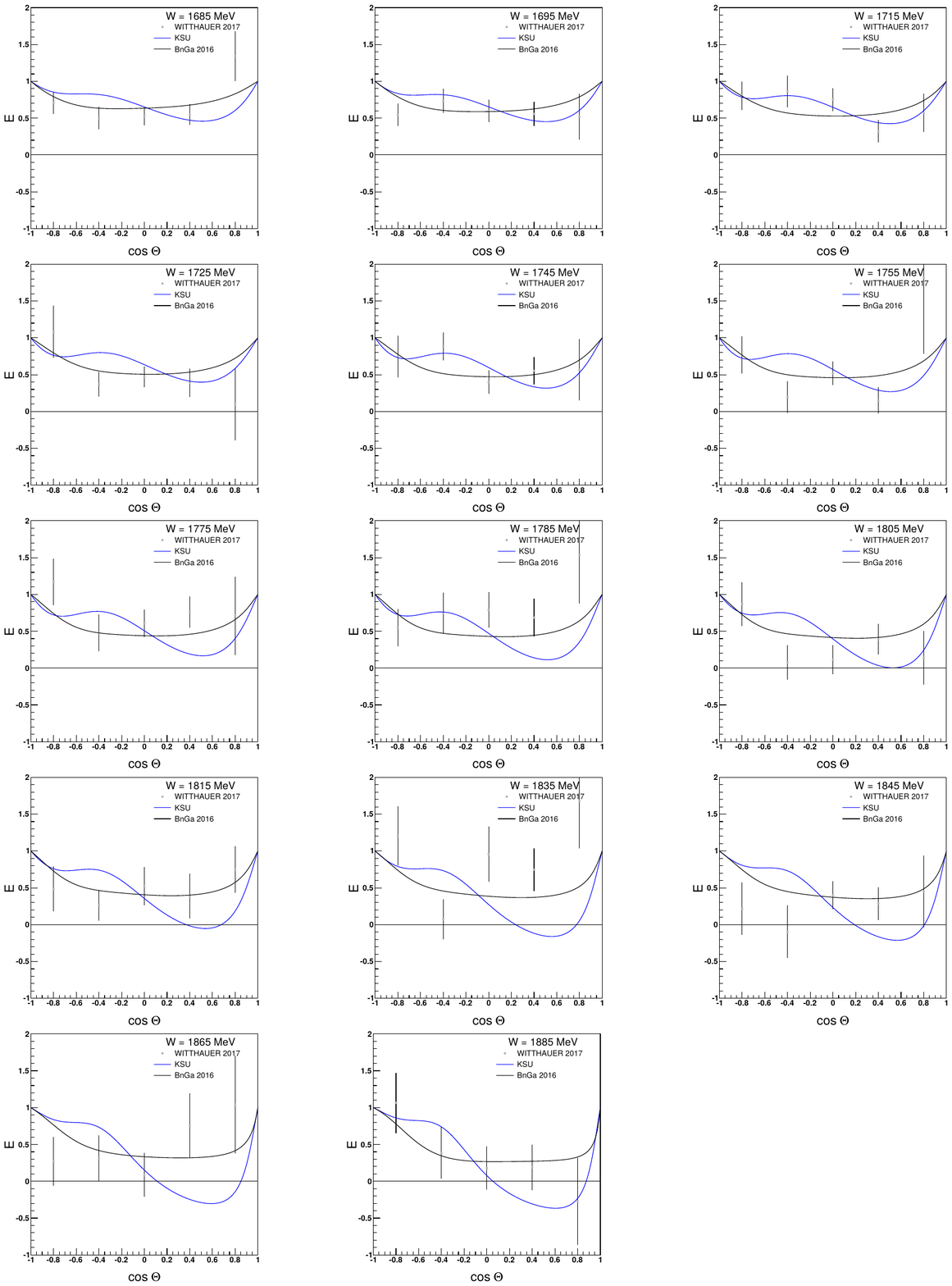}	
	\caption{Fits to $E$ data for $\gamma n \rightarrow \eta n$ at $W$ = 1685 to 1885~MeV. See text for references.}
\end{figure}

\renewcommand{\ObsName}{T}
\renewcommand{\printObs}{$T$}
\noindent
\begin{figure}
	\includegraphics[scale=\scalefac,trim={18mm} {118mm} {\trimC} {13mm},clip=true]{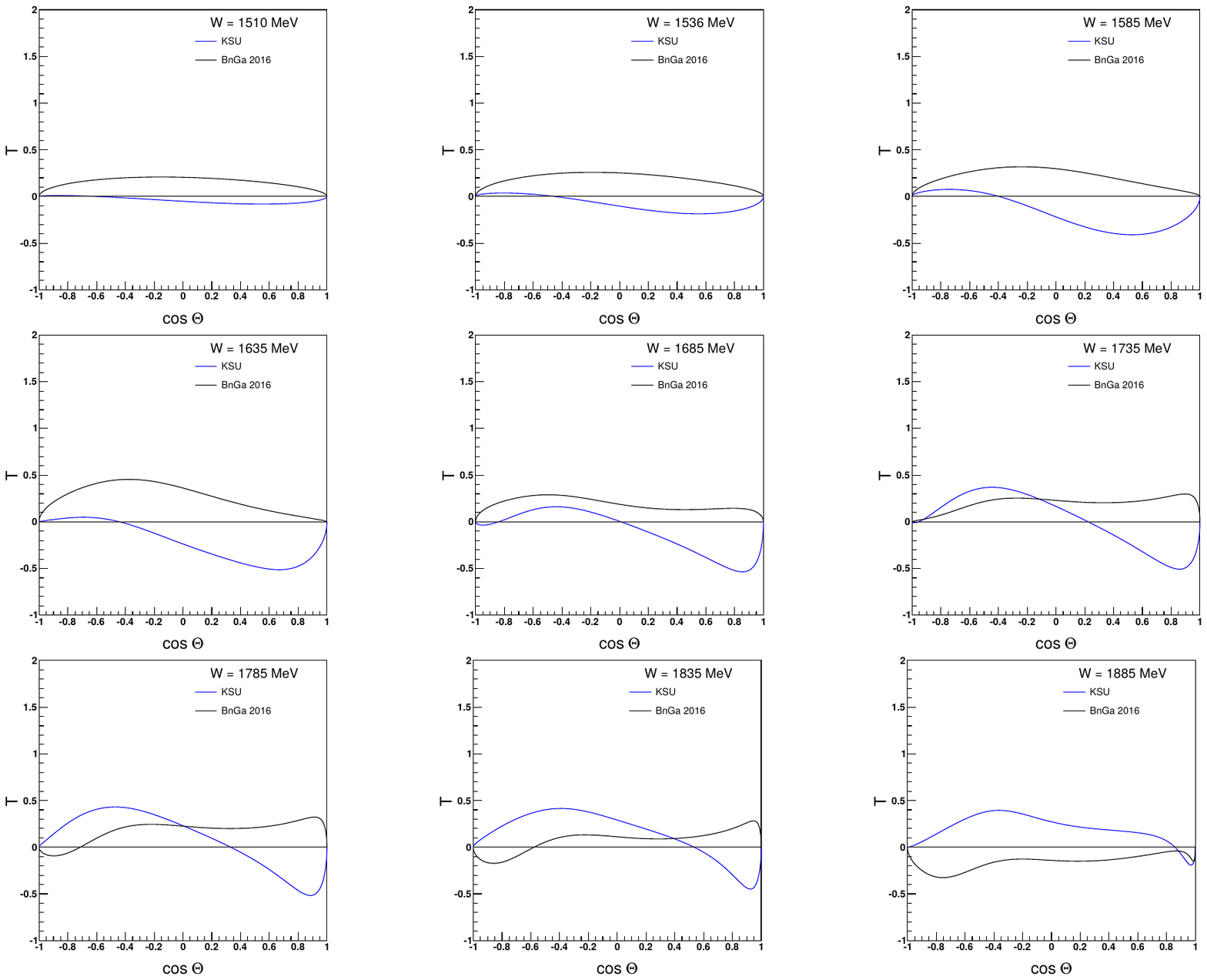}	
	\caption{Fits to preliminary $T$ data (not shown) \cite{Krusche15} for  $\gamma n \rightarrow \eta n$ at $W$ = 1510 to 1885~MeV.}
\end{figure}


\renewcommand{\ObsName}{F}
\renewcommand{\printObs}{$F$}
\begin{figure}
	\includegraphics[scale=\scalefac,trim={18mm} {118mm} {\trimC} {13mm},clip=true]{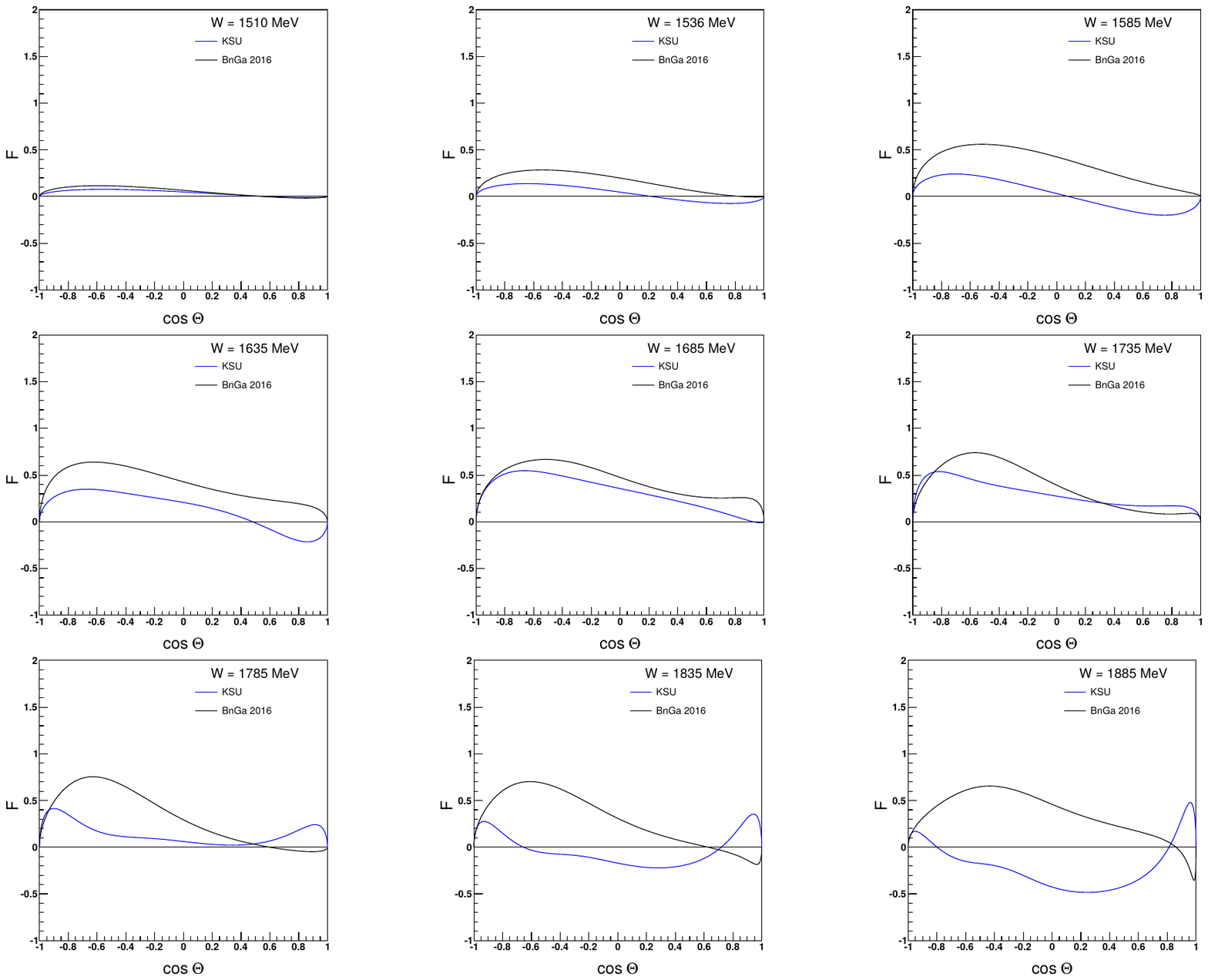}	
	\caption{\label{GPEP_Fig30}Fits to preliminary $F$ data (not shown) \cite{Krusche15} for $\gamma n \rightarrow \eta n$ at $W$ = 1510 to 1885~MeV.}
\end{figure}


\bibliography{basename of .bib file}

\end{document}